\pdfoutput=1
\documentclass[journal]{vgtc}                

\usepackage[utf8]{inputenc}

\ifpdf
  \pdfoutput=1\relax                   
  \usepackage{graphicx}                
  \DeclareGraphicsExtensions{.pdf,.png,.jpg,.jpeg} 
\else
  \usepackage{graphicx}                
  \DeclareGraphicsExtensions{.eps}     
\fi%

\graphicspath{{figures/}{pictures/}{images/}{./}} 

\usepackage{microtype}                 
\PassOptionsToPackage{warn}{textcomp}  
\usepackage{textcomp}                  
\usepackage{mathptmx}                  
\usepackage{times}                     
\usepackage{cite}                      
\usepackage{tabu}                      
\usepackage{booktabs}                  

\newcommand{\myParagrapho}[1]{\smallskip\noindent\textbf{#1}\xspace}

\onlineid{0}

\vgtccategory{Research}
\vgtcpapertype{algorithm/technique}

\usepackage{cellspace}

\usepackage{ifpdf}

\ifpdf
  \pdfoutput=1\relax                   
  \usepackage{graphicx}                
  \DeclareGraphicsExtensions{.pdf,.png,.jpg,.jpeg} 
\else
  \usepackage{graphicx}                
  \DeclareGraphicsExtensions{.eps}     
\fi%

\usepackage{tabularx}

\graphicspath{{figures/}{pictures/}{images/}{./}}

\usepackage{microtype}         
\PassOptionsToPackage{warn}{textcomp} 
\usepackage{textcomp}                 
\usepackage{mathptmx}                
\usepackage{times}                    
       
\usepackage{cite}                   
\usepackage{tabu}                   
\usepackage{booktabs}               
\usepackage[normalem]{ulem} 
\usepackage{xspace}
\usepackage{amsfonts}
\usepackage{amsmath}
\usepackage{multirow}  

\usepackage[utf8]{inputenc}
\usepackage[T1]{fontenc}
\usepackage{balance}
\usepackage[hyphens]{url}
\usepackage{tabularx}

\usepackage{wrapfig,epsf}

\usepackage{xcolor}

\usepackage[hidelinks]{hyperref}

\newcommand{\ie}{\emph{i.e.}\xspace}
\newcommand{\eg}{\emph{e.g.}\xspace}

\newcommand{\myparagraph}[1]{\smallskip\noindent\textbf{#1}\xspace}

\newcommand{\figref}[1]{\hyperref[#1]{\textbf{Figure~\ref*{#1}}}}
\newcommand{\figsref}[1]{\hyperref[#1]{Figures~\ref*{#1}}}
\newcommand{\tabref}[1]{\hyperref[#1]{Tab.~\ref*{#1}}}
\newcommand{\secref}[1]{\hyperref[#1]{Sec.~\ref*{#1}}}
\newcommand{\algref}[1]{\hyperref[#1]{Alg.~\ref{#1}}}

\newcommand{\blue}[1]{\textcolor{black}{#1}}

\title{LargeNetVis: Visual Exploration of Large Temporal Networks Based on Community Taxonomies}

\author{Claudio D. G. Linhares*, 
Jean R. Ponciano*, 
Diogenes S. Pedro, 
Luis E. C. Rocha, 
Agma J. M. Traina, 
and Jorge Poco}

\authorfooter{
\item[*] The first two authors contributed equally to this work.
\item
C. Linhares, D. Pedro and A. Traina are with Institute of Mathematics and Computer Sciences, University of S\~ao Paulo, S\~ao Carlos, Brazil. E-mails: \{claudiodgl, diogenes.pedro\}@usp.br, \{agma\}@icmc.usp.br.
\item
Luis E. C. Rocha is with Department of Economics \& Department of Physics and Astronomy, Ghent University, Belgium. E-mail: luis.rocha@ugent.be.
\item
J. Ponciano and J. Poco are with School of Applied Mathematics, Funda\c{c}\~ao Getulio Vargas, Rio de Janeiro, Brazil. E-mails: \{jean.ponciano, jorge.poco\}@fgv.br.
}

\abstract{
Temporal (or time-evolving) networks are commonly used to model complex systems and the evolution of their components throughout time.
Although these networks can be analyzed by different means, visual analytics stands out as an effective way for a pre-analysis before doing quantitative/statistical analyses to identify patterns, anomalies, and other behaviors in the data, thus leading to new insights and better decision-making.
However, the large number of nodes, edges, and/or timestamps in many real-world networks may lead to polluted layouts that make the analysis inefficient or even infeasible.
In this paper, we propose LargeNetVis, a web-based visual analytics system designed to assist in analyzing small and large temporal networks. It successfully achieves this goal by leveraging three taxonomies focused on network communities to guide the visual exploration process.
The system is composed of four interactive visual components: the first (Taxonomy Matrix) presents a summary of the network characteristics, the second (Global View) gives an overview of the network evolution, the third (a node-link diagram) enables community- and node-level structural analysis, and the fourth (a Temporal Activity Map -- TAM) shows the community- and node-level activity under a temporal perspective.
}

\vgtcinsertpkg

\teaser{
  \centering
  \includegraphics[width=\linewidth]{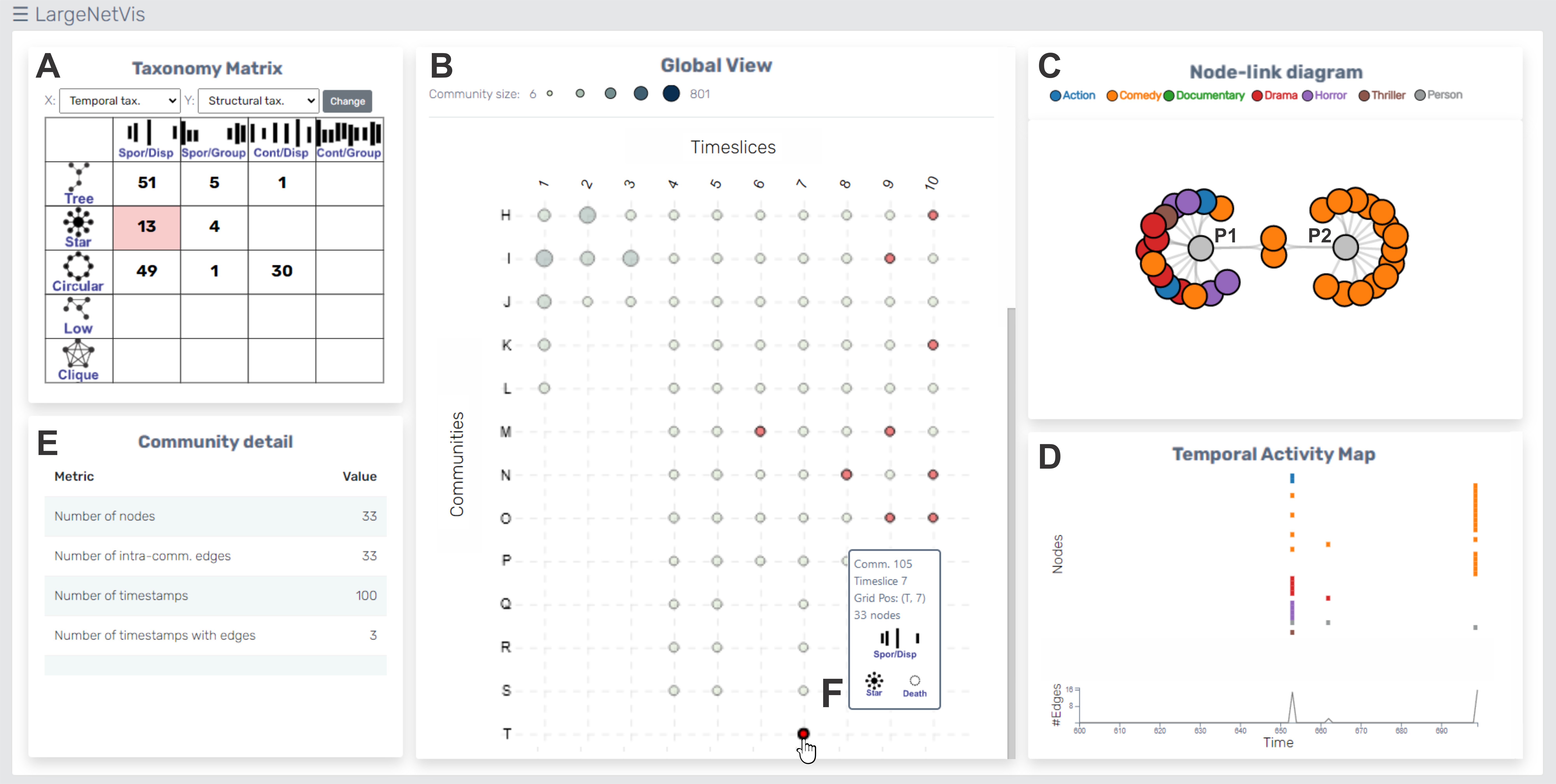}
  \vspace{-0.7cm}
  \caption{LargeNetVis is a web-based visual analytics system designed to support the visual exploration of temporal networks with up to a few thousand nodes and timestamps. The system is composed of four linked visual components. While the \textit{Taxonomy Matrix} (A) and \textit{Global View} (B) enable global-level analysis, the \textit{node-link diagram} (C) and the \textit{Temporal Activity Map} (D) enable structural and temporal local-level analysis, respectively. The system also provides a panel with the network, community, or node numerical information (\textit{Community detail} (E)). \blue{A tooltip with extra information is shown when a community is hovered over in Global View (F).}}

  \label{fig:teaser}
}

\begin{document}


\maketitle

\section{Introduction}
\label{sec:introduction}

Graphs are mathematical objects used to map the relation (edges) of different elements (nodes). Complex networks use graphs to model real-world events by studying the structure and dynamics of complex systems~\cite{complexNetworks}. Moreover, when the dynamics of these networks are mapped through the time information, they are called temporal (or time-evolving) networks~\cite{Lehmann2019}. Information visualization techniques help in analyzing temporal networks as they allow exploratory analysis and also enhance the identification of structural and temporal patterns that involve individuals and groups of elements. Visualization layouts have been proposed in the last decades for these purposes, from timeline-based approaches and animations to hybrid visualizations~\cite{SurveyDynamicVisualization}.

In our context, visual scalability refers to the ability of a layout to work properly and adapt to small and large networks. This capability is particularly important considering the number of real-world networks regularly created with the acquisition of big data from different application domains, including online social media, biological, and others.
A crucial definition is, therefore, \textit{what a large network is}. Such definition depends on the type of study and the number of elements in the network. Non-temporal networks with millions of nodes and edges are commonly analyzed in statistical studies, and these networks are often referred to as large networks~\cite{statisticsLarge}.
For visualization purposes, on the other hand, this definition may vary according to the task and aspects such as edge density and visual clutter. To exemplify how abstract this definition is in this context, while Mi et al.~\cite{informatics3040023} consider large (non-temporal) networks as those with millions of nodes, a more recent survey evaluated the size of networks in 124 visualization papers and classified networks with more than 200 nodes as very large ones~\cite{YOGHOURDJIAN2018264}.

We consider in this work large temporal networks as those with a few thousand nodes or timestamps. This decision was made based on the trade-off between the spread of nodes and edges throughout time, which may reduce the number of elements simultaneously shown depending on the layout (especially when adopting timeslicing~\cite{PONCIANO2021170}), and the increased complexity that the temporal dimension brings to timeline-based and animated visualizations (\eg, mental map preservation). Although layouts and visualization systems designed to handle temporal networks work well with relatively small networks, most of them are ineffective used when dealing with large ones. The main reason for that is the high level of visual clutter in the layouts, which occurs depending on the network density and dynamics and is mainly caused by overlaps of nodes, edges, or groups~\cite{SurveyDynamicVisualization}. Layouts that suffer from this issue include the traditional node-link diagram (animated or not), alluvial diagrams~\cite{Rosvall2010}, \blue{GraphFlow~\cite{GraphFlow}}, and the \textit{Massive Sequence View} (MSV) representation~\cite{Elzen2014}. 
Since each layout can benefit from a set of features, different methods have been proposed to improve layout readability and thus lead to new insights into the data. 
Examples include methods focused on node positioning~\cite{Elzen2014}, node and edge sampling~\cite{sampleStream1,EdgeSampling2018}, timeslicing~\cite{PONCIANO2021170,wang}, and representation simplification~\cite{Stanley2018,8249874, 101137}. Even considering recent advances in layouts, algorithms for clutter reduction, and interactive visualization systems, enabling compelling visual exploration of large temporal networks remains a challenging and open problem~\cite{SurveyDynamicVisualization}.

\blue{Meaningful groups of nodes are frequently observed in real-world temporal networks, for example, those composed of nodes that interact more often between themselves than with nodes from other groups, the so-called communities~\cite{Fortunato2016}. Communities can represent, \eg, groups of friends, co-authors, brain regions, similar proteins in a biological network, and others.}
In this paper, we propose LargeNetVis, a web-based visual analytics system created to assist experts and practitioners in analyzing small and large temporal networks. \blue{The system relies on network community detection to guide users in finding and inspecting regions of interest that may contain potentially relevant network elements.}
More specifically, LargeNetVis leverages three community taxonomies to allow users to understand these communities' structural, temporal, and evolutionary patterns. Furthermore, users can explore communities with characteristics of interest and analyze their elements' behaviors through the four linked views provided by the system. 

In summary, our main contributions are: i) Integration of three taxonomies for communities in temporal networks, each accounting for a particular and relevant type of pattern; ii) LargeNetVis, a visual analytics system with multiple linked views that leverages the mentioned taxonomies to enhance the exploration of large temporal networks\,---\,the usefulness and effectiveness of the system were validated through a user study with 14 participants not involved in the system's development; and iii) Two usage scenarios showing relevant and non-trivial patterns found in large and real-world networks.

\section{Related work}

\blue{Different strategies for modeling temporal changes and identifying patterns can be employed, \eg, with the use of graphlets and motifs~\cite{CLEMENTI201519, 1081893, btv227}.} In this study, we leverage the temporal network community structure to identify patterns and enhance the visual analysis. This section discusses existing community taxonomies, aspects related to the visualization, and available systems.

\subsection{Network community taxonomies}
\label{network_community_taxonomy}

\blue{Different taxonomies have been proposed for network visualization. They categorize the network elements according to different criteria, \eg, clusters or dimensions~\cite{102312,103389}, thus optimizing the process of finding elements with characteristics of interest. LargeNetVis leverages three taxonomies focused on different aspects of community categorization\,---\,the community's structural, temporal, and evolutionary behaviors\,---\,to guide the network exploration. The categories considered by these taxonomies have also been analyzed by previous works~\cite{Li2017, 3441301, 2466444, CNO, Rosvall2010,PEREIRA2021100904, 8930777} and represent the core of our approach.}

\begin{figure}[t]
	\includegraphics[width=\linewidth]{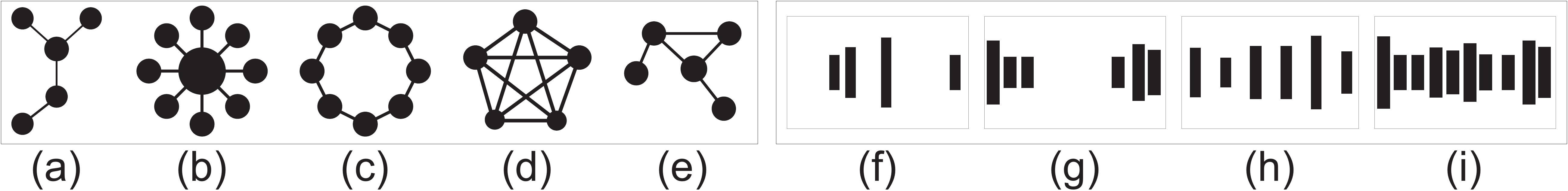}
	\centering
	\vspace{-0.5cm}
	\caption{\blue{Structural (a-e) and temporal (f-i) taxonomies. Vertical bars in the temporal taxonomy represent edges (y-axis) over time (x-axis).}}
	\label{fig:structure}
	\vspace{-0.5cm}
\end{figure}

\looseness=-1
\myParagrapho{Structural.} A useful manner of understanding a network's overall topology is categorizing its communities according to their structural patterns, such as stars, cliques, circulars, and others. For example, stars (see definition below) are commonly found on social networks. Different names for the same categories have been used in previous studies, such as clique and full, star and egocentric, loop and circular, among others~\cite{Li2017,3441301}. Based on Chenhui et al.'s work~\cite{Li2017}, we use five types of community structure classification: Tree, Star, Circular, Clique, and Low-connectivity (Fig.~\ref{fig:structure}(a-e)). 
The \textit{Tree} topology (Fig.~\ref{fig:structure}(a)) has no cycles and differs from the \textit{Star} topology (Fig.~\ref{fig:structure}(b)), which  corresponds to a central node concentrating most of the connections, with the peripheral/children nodes usually with only one connection. In the \textit{Circular} case (Fig.~\ref{fig:structure}(c)), every node has only two connections. The \textit{Clique} (Fig.~\ref{fig:structure}(d)) represents a case where all nodes are connected with each other (complete graph). Cases where none of the previous categories are suitable are called \textit{Low-connectivity} ones (Fig.~\ref{fig:structure}(e)). In real-world networks, a single community may contain more than one type of structure (details in Sec.~\ref{limitation}) or be formed by variations of these types (\eg, a circular formation with some additional edges linking non-adjacent nodes). By similarity, we categorize communities from this latter case as their reference structure (as circular in the previous example, for instance).

\myParagrapho{Temporal.} \blue{Another important aspect concerns the overall temporal behavior of a community's internal edges (intra-community edges), including their temporal activity (\ie, how frequent they occur over time) and distribution (\ie, how dispersed they are from each other).} A
previous study focused on a classification suitable for these temporal
patterns, called Temporal taxonomy~\cite{CNO}. The first part of this taxonomy is the frequency, \ie, if the intra-community edges are so frequent over time that the community is categorized as having \textit{Continuous} activity, or if there is occasional activity and the edges occur in a \textit{Sporadic} fashion. The taxonomy also considers dispersion information, \ie, if the edges of consecutive timestamps are close to each other (\textit{Grouped}) or if they are sparse and, consequently, \textit{Dispersed}.
By combining frequency and dispersion, this taxonomy creates a temporal classification that comprises four combinations, as illustrated in Fig.~\ref{fig:structure}(f-i). In the first two combinations (Fig.~\ref{fig:structure}(f-g)), communities are categorized as Sporadic and Dispersed (Fig.~\ref{fig:structure}(f)), \ie, there are no dense regions and at least one timestamp has no edges, or Sporadic and Grouped (Fig.~\ref{fig:structure}(g)), when all edges are in a dense region but not continuous over time (we can think of as peaks of seasonal events). On the other hand, edges may occur frequently enough to be considered Continuous, and they can be grouped (Fig.~\ref{fig:structure}(i)) or not (Fig.~\ref{fig:structure}(h)).

\myParagrapho{Evolution.} The final aspect that we consider is the network community evolution, which is essential to understand the dynamics and lifecycle of these groups of nodes, with events such as birth and growth. We consider the community taxonomy proposed by Pereira et al.~\cite{PEREIRA2021100904}, which includes six categories of evolutionary events: Birth, Death, Grow, Contract, Split and Merge. The first appearance of a community is classified as \textit{Birth}, while disappearance is considered as \textit{Death}. A community may maintain most of its nodes and \textit{Grow} with the addition of new ones or \textit{Contract} with the loss of nodes. A community may also \textit{Split} into two new ones or \textit{Merge}, leading to a new community.

\subsection{Visualization of temporal networks}

For temporal network visualization, layouts may be divided into timeline-based and animation techniques~\cite{SurveyDynamicVisualization}. While animation suffers from mental map preservation in timestamps with high variation in the number of elements shown, timeline-based layouts usually lack screen space for a high number of nodes and timestamps. 
Among the animated layouts, the node-link diagram is still one of the most preferred representations~\cite{SurveyDynamicVisualization}. \blue{It is used, \eg, to visualize online networks, where nodes and edges arrive in a real-time manner~\cite{9231268}.}
While it can suffer from scalability limitations when applied to \blue{networks with several nodes and edges per timestamp}, representation simplification (\eg, super nodes~\cite{Stanley2018}) may attenuate this problem.
Despite strengths and limitations, both timeline and animations are efficient for temporal tasks. Timeline representations, however, are better for tasks that depend on the analysis of more than two timestamps~\cite{SurveyDynamicVisualization}. 

A few studies try to enable visual analyses of non-temporal networks with millions of nodes~\cite{informatics3040023, Lin2013DemonstratingIM}. These studies, however, have focused only on general rather than local structures. Their visualizations lack user validation and are not suitable for temporal networks. In the context of temporal hypergraphs, \ie, temporal networks where a single edge (called hyperedge) can connect more than two nodes, Valdivia et al.~\cite{PaohVis} proposed a timeline-based visualization that contains an aggregation of hyperedges and was tested in hypergraphs with up to 500 nodes. Also focusing on temporal hypergraphs, Fischer et al.~\cite{9222341} proposed a matrix-based visualization containing filtering, ordering, and interaction techniques. Temporal hypergraphs are significantly different from regular temporal networks, especially regarding structural patterns. Our focus is on the latter type of network.

\blue{Some studies combine temporal and structural analysis with summarization methods without considering community taxonomies~\cite{8249874}.} The use of a network community taxonomy to summarize the visualization was initially proposed by Li et al.~\cite{Li2017}. They focused on non-temporal networks and used the Structural taxonomy described in Sec.~\ref{network_community_taxonomy}. Later, Wang et al.~\cite{3441301} extended this approach to temporal networks. They proposed a pipeline to group nodes according to common characteristics (\eg, via community detection), applying optional sampling, classifying the groups according to the Structural taxonomy, and providing a system with multiple views containing animation-based visualization, network statistics, and parallel coordinates. While Wang et al. demonstrated the importance of using a network community taxonomy to simplify temporal network visualization, they focused only on the structural aspect of the network and did not validate their pipeline and system with a user study.

\blue{Timeline-based techniques can use timeslices to represent the network time dimension. In our context, a timeslice is defined as a short observation period (\eg, 1-hour or 1-day length) and comprises all nodes and edges that occur in that period. Note that each timeslice includes several timestamps of the network. 
Although timeslicing methods are commonly used with discrete timestamps, they fail to reliably represent networks with continuous real-valued time  (the so-called continuous (or event-based) dynamic networks)~\cite{HOLME201297,DynNoSlice2, DynNoSlice}.
A set of visualization techniques have been proposed for this type of network, such as the DynNoSlice~\cite{DynNoSlice2, DynNoSlice}, which employs space-time cubes to visualize the network's events, and MultiDynNoS~\cite{MultiDynNoS}, which presents a multilevel approach to reduce running time when visualizing these networks. LargeNetVis is designed for networks with discrete timestamps. }

\subsection{Comparison with network visualization systems}

We compared different systems for network visualization using the corresponding papers' descriptions, system images, case studies, comprehensive surveys (\eg,~\cite{SurveyDynamicVisualization, Beck2014, SurveyGroupStructures}), and external videos, when available. Related to network visualization, there is a variety of systems and papers that propose techniques to enhance pattern identification~\cite{Pregel, GraphChi, NetworkX, Cytoscape, DyNetViewer, Gephi, 9222341, timeArcs, 6065001, Elzen2014, nr-aaai15, pajek, DyEgoVis, tulip}, in which we applied a selection criterion to compare. We filtered in systems that have animated or timeline-based representations, high relevance and similarity with our system, and a high number of paper citations and/or downloads. Matching the previous criteria, we selected six systems.

\begin{table}[t]
\centering
\caption{Comparison of network visualization systems according to: (1) grouping; (2) sampling; (3) \blue{animation}; (4) timeline; (5) overview; (6) local view; (7) visual scalability; (8) \blue{multiple views}.}
\label{tab:comparison}
\resizebox{0.48\textwidth}{!}{%
\begin{tabular}{lllllllll}
\toprule
                  & 1 & 2 & 3 & 4  & 5 & 6 & 7 & 8  \\
\midrule
Gephi~\cite{Gephi}     & $\checkmark$  & $\checkmark$  & $\checkmark$  &      & $\checkmark$  & $\checkmark$  & $\checkmark$  &    \\
Pajek~\cite{pajek}          & $\checkmark$  &   & $\checkmark$  &   &    $\checkmark$  &  $\checkmark$ & $\checkmark$  &    \\
DyNetViewer~\cite{DyNetViewer}           & $\checkmark$  & $\checkmark$  & $\checkmark$  &   &   $\checkmark$  & $\checkmark$  &   &    \\
DyNetVis~\cite{dynetvisSac}   & $\checkmark$  &  & $\checkmark$   & $\checkmark$  &    $\checkmark$  &  $\checkmark$ & $\checkmark$  &    \\
\blue{DyEgoVis~\cite{DyEgoVis}} & $\checkmark$ & & & $\checkmark$ & $\checkmark$ & $\checkmark$ & & $\checkmark$ \\
Wang et al.'s~\cite{3441301}       & $\checkmark$  &  $\checkmark$  & $\checkmark$ &     & $\checkmark$  &   & $\checkmark$  & $\checkmark$   \\
LargeNetVis (Ours)   &  $\checkmark$ &  $\checkmark$ &  &  $\checkmark$ &  $\checkmark$  & $\checkmark$  &  $\checkmark$ & $\checkmark$   \\
\bottomrule
\end{tabular}%
}
\vspace{-0.3cm}
\end{table}

Table~\ref{tab:comparison} summarizes seven systems considering 8 relevant aspects inspired by~\cite{6415893}. The \textit{grouping} category (1) is related to node clustering or community detection, and \textit{sampling} (2) is whether the system provides node, edge, or group sampling.
The temporal dimension can be considered through \textit{animation} (3) and/or \textit{timeline} (4). We focus on the latter due to its advantage in temporal tasks that depend on more than two timestamps~\cite{SurveyDynamicVisualization}. \blue{Likewise, a system may enable analysis at a global scale (\textit{overview}) (5) and/or offer views dedicated to analyzing particular groups of nodes or time intervals of interest (\textit{local view}) (6).
We consider that a system offers \textit{visual scalability} (7) if it works for small and large networks\,---\,we considered the existing case and user studies to assess this criterion. Finally, the \textit{multiple views} category (8) refers to whether the system provides different views on a single screen.}
Notably, DyNetVis~\cite{dynetvisSac} is the only system among the compared ones that provide both timeline and animations, \blue{while only Wang et al.'s~\cite{3441301}, DyEgoVis~\cite{DyEgoVis} and ours have multiple views; these three are also the only web-based systems in the table}. At last, the only system from Table~\ref{tab:comparison} validated through a user study is ours (Sec.~\ref{ux}).

\section{Design Tasks}

\looseness=-1
Different task taxonomies have been proposed to guide the design and evaluation of layouts and systems for temporal network visualization~\cite{mainTaskTaxonomy1,mainTaskTaxonomy2,taskTaxonomy3,taskTaxonomy4}. As we are interested in enabling the analysis of structural and time-evolving patterns at both global and local granularity levels (\ie, network, groups, and nodes), we designed our views and interactions to meet tasks considering the taxonomies proposed by Bach et al.~\cite{mainTaskTaxonomy1} (taxonomy~1), and by Ahn et al.~\cite{mainTaskTaxonomy2} (taxonomy~2).
Taxonomy~1 considers three categories of low-level tasks, named \textit{temporal} (when), \textit{topological} (where), and \textit{behavioral} (what), that may be combined to obtain compound tasks. According to this taxonomy, a system must have the following characteristics to support tasks from these categories~\cite{mainTaskTaxonomy1}: 

\myparagraph{T1:} allow easy identification and reaching of specific time steps (when); 

\myparagraph{T2:} allow easy identification, situation, and tracking of elements (\eg, nodes or groups) with particular properties (where); 

\myparagraph{T3:} allow the understanding of the nature of changes that affect the elements (what).

\begin{figure*}[t]
	\includegraphics[width=0.9\linewidth]{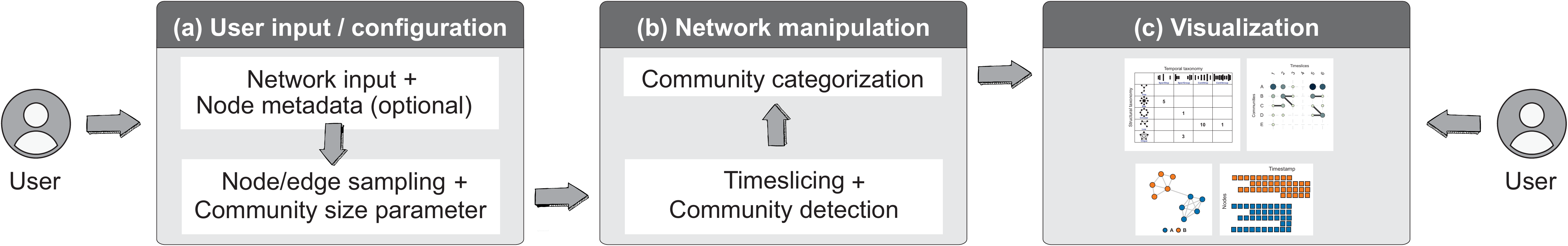}
	\centering
	\vspace{-0.3cm}
	\caption{LargeNetVis's workflow. (a) The user provides a network of interest and the corresponding node metadata (optional). He/She can also apply node/edge sampling and adjust the minimum community size parametrization. (b) The system divides the network into timeslices of equal length and then detects network communities inside each of them. These communities are then categorized according to three taxonomies from the literature. (c) Finally, users explore communities with characteristics of interest and their nodes and edges through the four linked visual components.}
	\label{fig:workflow}
	\vspace{-0.3cm}
\end{figure*}

Focused on aspects related to the network evolution, taxonomy~2 describes tasks related to the three dimensions \textit{entity} (an element of interest, such as node, edge, group, network), \textit{property} (a characteristic of an element, given by centrality measures or domain properties, for example), and \textit{temporal feature} (an event that causes changes in the state of an element, such as growth or death). Although all three dimensions are somehow related to categories also foreseen by taxonomy~1, the last dimension provides more details about existing types of temporal events, four of which are also observed in the community evolution taxonomy described in Sec.~\ref{network_community_taxonomy} (birth/death and growth/contraction). Since the community evolution taxonomy considers merge/split events in addition to the four mentioned types, we restricted our task T3 to consider all six types of events from this taxonomy. Finally, our system was also designed to support:

\myparagraph{T4:} identification of two other types of events from taxonomy~2's third dimension: single occurrences (\ie, the addition or deletion of an entity) and \textit{peak or valley} events (\ie, abrupt increases or decreases of an element property).

\section{LargeNetVis}
\label{LargeNetVis}

This section describes LargeNetVis's workflow and visual components, which were developed to support our design tasks as depicted in Table~\ref{tab:views_tasks}.

\subsection{Workflow}
\label{workflow}

Fig.~\ref{fig:workflow} describes the system's workflow. First, users provide a temporal network of interest and its respective node metadata (optional). Then, they can apply one of the node and edge sampling methods offered by the system, including random node/edge sampling and Snowball sampling~\cite{snowballSampling}, a method suitable for large networks. Other methods, \eg, EOD~\cite{EdgeSampling2018}, could be easily incorporated into the system. 
Users are also free to report the minimum number of nodes that a network community must have in order not to be ignored by the visualization\,---\,the community detection procedure is a further step that will be explained later on. Very small communities are often perceived in large networks, and they can pollute the layout with irrelevant information.

The next step is to divide the network data according to a user-defined number of timeslices (Fig.~\ref{fig:workflow}(b)).
\blue{To assist users in choosing a proper number of timeslices for analysis, LargeNetVis runs the nonuniform timeslicing method proposed by Ponciano et al.~\cite{PONCIANO2021170} with all edges having the same importance. By considering local edge density, the method defines the best timeslice length for each sliding window of timestamps. LargeNetVis then suggests a range of potentially suitable numbers of timeslices based on how many timeslices are required to meet from the minimum to the maximum length returned by Ponciano et al.'s algorithm. Within this range, the default value meets the average timeslice length, as recommended for global pattern identification~\cite{PONCIANO2021170}. Note that the choice of the number of timeslices strongly affects the quality of results and response time of the remaining workflow steps, consequently affecting the visualization and pattern identification.}

For each timeslice, network communities are detected using the \textit{Instant Optimal} strategy~\cite{3172867}, which prioritizes the quality of the community structure rather than temporal consistency. In this strategy, communities are detected separately for each timeslice and the overlap between communities from consecutive timeslices is computed. Any non-overlapping community detection algorithm could be used in this step. 
\blue{The LargeNetVis current version employs Louvain~\cite{louvain}, a greedy and agglomerative algorithm based on modularity optimization that finds the best network partition by establishing groups (communities) with densely connected nodes. Although modularity optimization faces the issue of resolution limit, which may lead small and meaningful communities to be clustered into larger ones~\cite{resolutionLimit,traag2019louvain}, Louvain is still one of the most popular and recommended algorithms~\cite{Fortunato2016}. Also, its computational complexity is linear on the number of edges, which is an essential advantage in our case. We point the reader to the survey by Fortunato and Hric~\cite{Fortunato2016} for a broader understanding of network community detection algorithms. Other algorithms (\eg, Infomap~\cite{infomap}) could be incorporated into the system with little effort. Different algorithms would return different communities; the detection quality directly affects the remaining steps of our workflow and, ultimately, the visualization and pattern identification.}

The system then categorizes each community of each timeslice using the three taxonomies described in Sec.~\ref{network_community_taxonomy} (Structural, Temporal, and Evolution). Finally, users visually explore the network data under different perspectives and granularity levels through the four linked visual components offered by the system (Fig.~\ref{fig:workflow}(c)). Respecting the well-established visual information-seeking mantra \textit{``overview first, zoom and filter, then details-on-demand''}~\cite{visualmantra}, these four components have different roles in the network analysis. Two of them provide an overview of the network structure and evolution (Sec.~\ref{systemOverview}), and two enable local-level analysis (Sec.~\ref{systemLocalView}). In addition to the four mentioned components, the system also shows a useful numerical information panel with network, community, or node information, depending on the granularity level of analysis. By default, this panel is named \textit{Network detail} panel, and it shows the overall number of nodes, edges, timestamps, and communities detected, as well as the mean modularity value returned by the detection procedure.

\begin{table}[t]
  \footnotesize
  \centering
  \caption{Visual components of LargeNetVis and the four tasks.}
  
  \label{tab:views_tasks}
  \resizebox{0.36\textwidth}{!}{%
  \begin{tabular}{lllllllll}
  \toprule
                                & T1 & T2 & T3 & T4 \\ 
    \midrule
    Taxonomy Matrix                &    & \checkmark     &  \checkmark   &         \\ 
    Global View                & \checkmark   &    \checkmark     & \checkmark       &        \\ 
    Node-link diagram                &    &  \checkmark  &           &   \\ 
    Temporal Activity Map                & \checkmark   &  \checkmark  &             & \checkmark \\ 
    \bottomrule
  \end{tabular}
  }
  \vspace{-0.3cm}
\end{table}

\subsection{Summary views}
\label{systemOverview}

\begin{figure}[t]
	\includegraphics[width=1\linewidth]{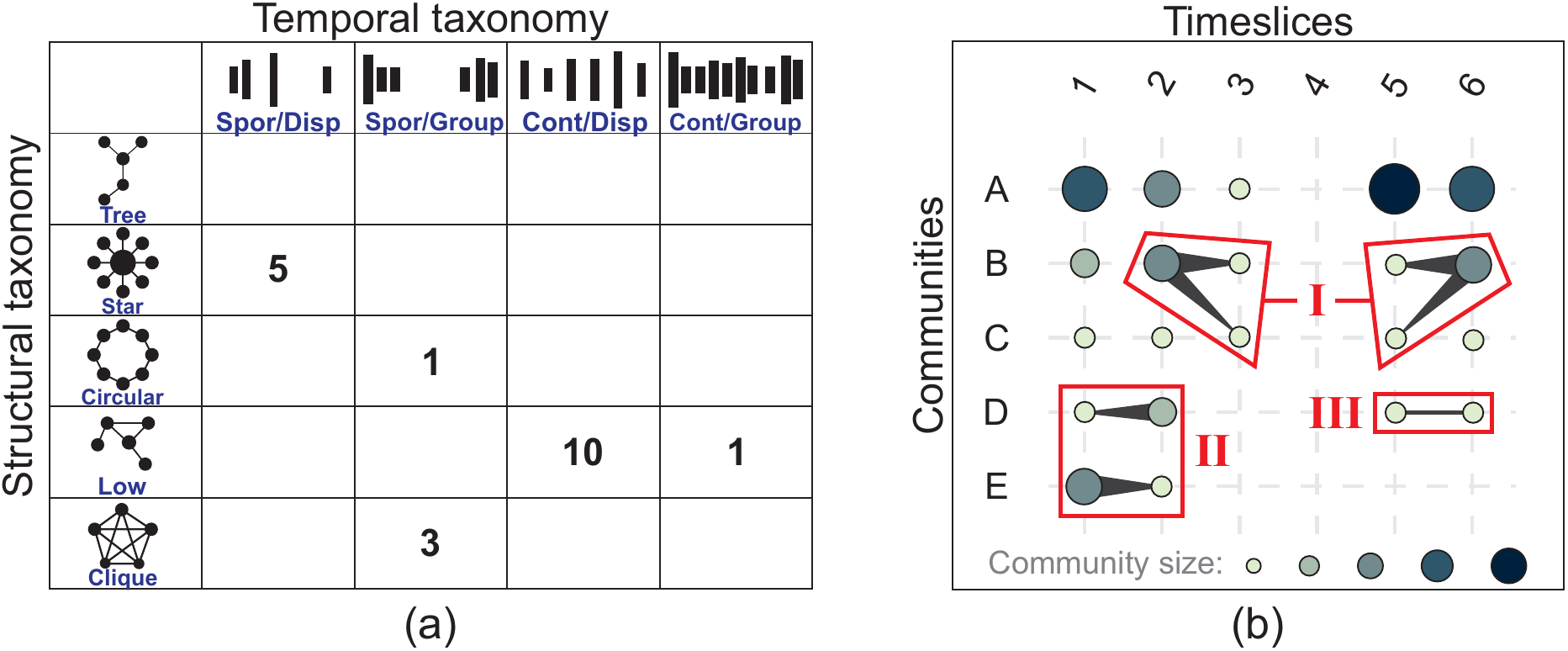}
	\centering
	\vspace{-0.6cm}
	\caption{Visual components of LargeNetVis that enable global analysis: (a) Taxonomy Matrix; (b) Global View. \blue{The evolutionary behaviors are: (I) split/merge; (II) grow/contract; and (III) preserve.} }
	\label{fig:didatic_overview}
	\vspace{-0.6cm}
\end{figure}

\myParagrapho{Taxonomy Matrix.} This view summarizes the general network structure, temporal distribution, and evolution by showing the number of communities that correspond to each combination of categories from the two user-selected taxonomies (tasks T2, T3). Users can freely combine taxonomies two-by-two, including a taxonomy with itself\,---\,in this case, the main diagonal would show how the network is characterized according to that taxonomy alone.
Fig.~\ref{fig:didatic_overview}(a) illustrates this view: in the example, three communities are \textit{Cliques} (structural taxonomy) and  \textit{Sporadic and Grouped} (temporal taxonomy).

Once the user has chosen two taxonomies for analysis, he/she can select a particular category combination of interest by clicking on its corresponding matrix cell. It is also possible to select multiple combinations, a row or a column, and even the whole matrix (by clicking on the top-left cell). 

\blue{An \textit{alternative} that we have considered for this view would be the adoption of bar charts depicting the number of communities that meet each category combination. The high number of possible combinations (there would be 30 bars when considering structural and evolution taxonomies), allied to the fact that the bar chart would emphasize the most frequent combinations while one may be interested in analyzing the less frequent ones, led us to adopt the described Taxonomy Matrix.}

\myParagrapho{Global View.} 
The communities that belong to the selected cell(s) in the Taxonomy Matrix are highlighted in Global View, a view focused on the community evolution behavior (Fig.~\ref{fig:didatic_overview}(b)). This way, users can locate communities with characteristics of interest (T2) and the timeslices they occur (T1), and also know the type of temporal changes affecting them (T3). Global View can be thought of as a grid where columns represent timeslices and rows represent communities, which are depicted as circles with varying colors and sizes, visual attributes that can map information of interest (for example, the community size or average node degree). The current version of LargeNetVis adopts redundant coding~\cite{info_visualization}, mapping the community size on both color and circle's size (Fig.~\ref{fig:didatic_overview}(b)).

As illustrated in Fig.~\ref{fig:didatic_overview}(b), Global View enables analyses that rely on the temporal distribution of communities, including the identification of idle periods (see timeslice \#4) and periods with many or few communities (compare timeslices \#1 and \#3, for instance). \blue{The evolutionary behaviors are visually represented as links with varying thickness between a ``from'' community (timeslice \#X) and a ``to'' community (timeslice \#(X+1)). The design of the links enables quick identification of splits and merges (Fig.~\ref{fig:didatic_overview}(b-I)), growth and contraction (Fig.~\ref{fig:didatic_overview}(b-II)), and preservation (Fig.~\ref{fig:didatic_overview}(b-III)), which occurs when the number of nodes in a community does not change (T3). Linking communities from different timeslices improves mental map preservation.} 

\blue{To avoid long links and overlaps, LargeNetVis implements a simple but efficient strategy that minimizes the link length. For the first timeslice (grid column), communities are positioned following the order they appear, \ie, the order they are returned by the community detection algorithm. For each remaining timeslice, we position ``to'' communities such that the Euclidean distance between them and their corresponding ``from'' communities is minimized; communities not linked so far go to the remaining positions in the order they appear.
In the end, we obtain minimal lengths for all links but merge-related diagonals (cases where there are two ``from'' for a single ``to''). To solve these cases, the ``from'' community that is far from its ``to'' (diagonal link) switches places with the non-linked community closest to the other ``from''. We provide the algorithm in the supp. material (Sec.~A). Note that other positioning strategies could have been adopted. The impact of different strategies in the layout is discussed in Sec.~\ref{limitation}.}

Users can zoom in/out and pan to explore particular regions of interest. When a circle is hovered over, the system shows a tooltip informing the timeslice number, the number of nodes in that community, the community position in the grid, and the three taxonomy categories that characterize that community (see Fig.~\ref{fig:teaser}(F)). When a link is hovered over, it is shown a tooltip with information about the communities and timeslices involved, the communities' sizes, and the type of event. Once the user has found a community of interest to analyze, he/she clicks on the corresponding circle and is redirected to the remaining two views (node-link diagram and Temporal Activity Map -- TAM), which will be described in the next section. Details of the clicked community, including its number of nodes, edges, and timestamps with edges, are also shown in the numerical information panel, which is now referred to as \textit{Community detail} panel (see Fig.~\ref{fig:teaser}(E)). If the user clicks on a circle that refers to a community categorized differently from those selected in the Taxonomy Matrix, the matrix cell representing this new community is automatically selected and the previously selected cells are deselected. Consequently, all communities in Global View that fit the newly selected categorization become highlighted (and only them). With this functionality, we help users to find communities that share the same properties as the ones they consider interesting. Note also that Global View enables exploratory visual analysis, \ie, users can go directly to this view without exploring Taxonomy Matrix first.

\looseness=-1
We have considered \textit{alternative design choices} for Global View. First, we adopted a timeline-based approach instead of an animation because timelines are better for tasks involving more than two time steps~\cite{SurveyDynamicVisualization}, \blue{an important advantage for identifying active/inactive time intervals (\eg, days and nights) (T1) and tracking a community of interest throughout the observation period (T2).}
After this decision, we considered the adoption of traditional alluvial (or Sankey) diagrams for this view. We discarded them after preliminary tests due to the amount of screen space they require for networks with many timeslices or communities per timeslice. Inspired by visualizations employed for node-level analysis~\cite{Elzen2014,PaohVis}, we created a compact representation that uses circles over time to indicate community activity and evolution. We also decided to map the community size using redundant code (circle's size and color) to enhance the identification of small/large communities \blue{(T2)}.

\subsection{Local views}
\label{systemLocalView}

\myParagrapho{Node-link diagram.}
After finding a community of interest, users may want to analyze it from a more detailed perspective, \eg, looking for intra-community nodes that share common properties (\eg, same metadata) or studying a node's interaction dynamics. LargeNetVis enables local-level analyses of these types (T2) through the traditional node-link diagram and a TAM~\cite{dynetvisSac}. Both views are simultaneously loaded when the user clicks on a circle in Global View.

\begin{figure}[t]
	\includegraphics[width=0.9\linewidth]{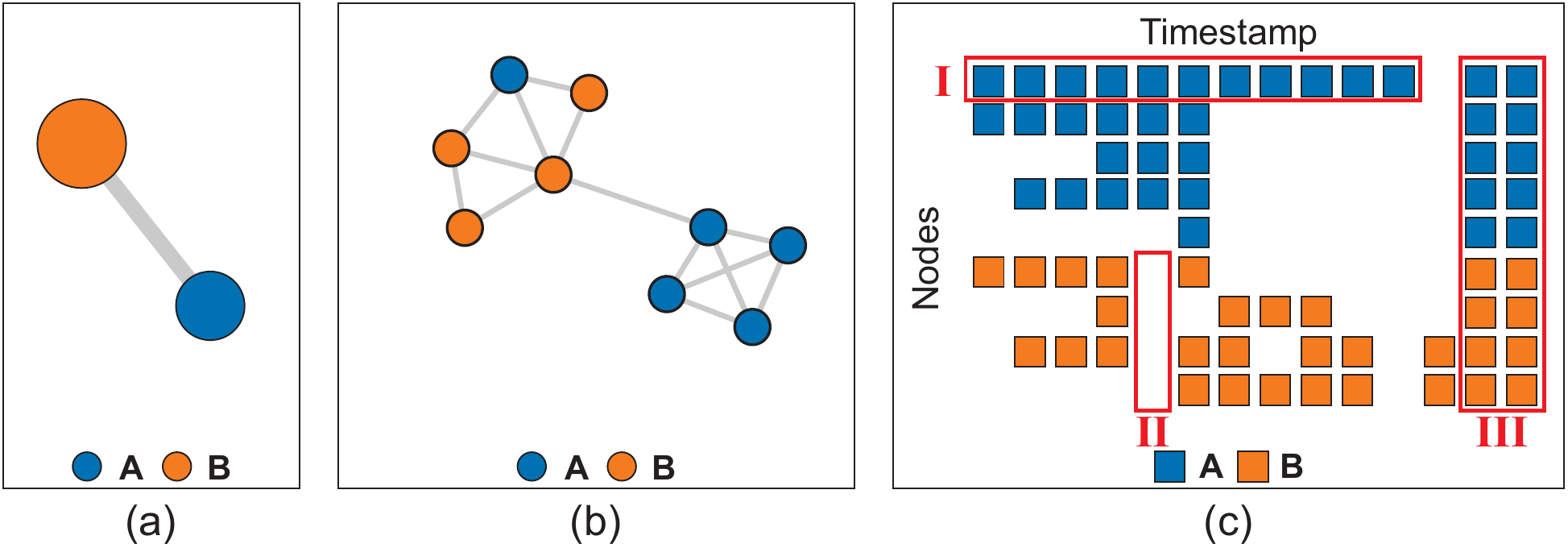}
	\centering
	\vspace{-0.4cm}
	\caption{\blue{Visual components of LargeNetVis that enable local analysis: (a) (Super)node-link diagram; (b) Node-link diagram; (c) TAM.}}
	\label{fig:didatic_local}
	\vspace{-0.5cm}
\end{figure}

Our node-link diagram, illustrated in Fig.~\ref{fig:didatic_local}(b), shows the intra-community nodes and edges in an aggregated fashion, \ie, by showing all of them at once regardless of the times of the edges. If categorical node metadata is provided (\eg, student's class or employee's department), the system enhances the analysis by associating different colors to nodes with different metadata values. Users are free to change any default color by using a color picker. The node positioning is defined such that nodes more frequently connected are close to each other (spring-force algorithm~\cite{ForceDirectedStructural}) \blue{and edge bundling is optionally applied to reduce clutter~\cite{bundle}. The impact of different positioning methods is discussed in Sec.~\ref{limitation}}. As occurs with Global View, users can zoom in, zoom out and pan for more detailed analysis or manual node repositioning. When a node (circle) is hovered over, the system shows a tooltip informing the node id and its metadata value (if any). If a circle is clicked, \blue{the corresponding node is highlighted in TAM (described below) and} the numerical information panel (now named \textit{Node detail} panel) exhibits this node's id, metadata value, normalized degree value, approximated and normalized betweenness centrality value (using 25\% of the nodes for computation), and closeness value. These centrality values are computed for the timeslice under analysis.

\blue{To enhance the analysis of large communities, the system also offers a multilevel node-link diagram for those communities with more than a user-defined number of nodes. In these cases, LargeNetVis shows, by default, a summarized representation where nodes (referred to as ``supernodes'') represent sub-communities and ``superedges'' indicate the presence of inter-sub-community edges (Fig.~\ref{fig:didatic_local}(a)). The algorithms used for ``supernode'' positioning and sub-community detection~\cite{CNO} are the same as before (spring-force and Louvain, respectively).
The color and size of a ``supernode'' (circle) map the predominant color among the sub-community's nodes and the number of nodes inside it, respectively. The ``superedge'' thickness maps the number of inter-sub-community edges between the two involved ``supernodes''. Users can freely navigate between summarized and non-summarized node-link diagrams. When a ``supernode'' is clicked, its nodes are highlighted in both non-summarized diagram and TAM.}

\blue{
We performed preliminary tests to compare our node-link diagram with an \textit{alternative approach} (the adjacency matrix visualization~\cite{matrix}). We chose the node-link diagram as it was easier to visualize the inherent structure of the communities (T2) using the diagram than with the matrix (recall Fig.~\ref{fig:structure}(a-e)).
}

\myParagrapho{Temporal Activity Map.} \blue{Besides the well-established node-link diagram, LargeNetVis includes the Temporal Activity Map (TAM), a layout successfully used for the visual exploration of networks from different domains~\cite{dynetvisSac}. TAM is a timeline-based visualization strategy focused on showing active nodes (\ie, connected to others) over time. The vertical and horizontal axes represent the nodes and timestamps, respectively, and a colored square is drawn whenever the corresponding node is active in the network. The edges are hidden in this layout. Our system exhibits TAM at the same time as the node-link diagram and uses the same colors for node metadata (T2). For coherence, colors changed through the mentioned color picker are also reflected in this layout.
As illustrated in Fig.~\ref{fig:didatic_local}(c), active nodes are represented by fixed-size squares that form ``blocks'' in the visualization that enable the identification of patterns related to active/inactive nodes and groups over time as well as intervals with or without activity (T1, T4). Some examples of patterns are shown in Fig.~\ref{fig:didatic_local}(c): the
presence of a node active in several consecutive timestamps (Fig.~\ref{fig:didatic_local}(c-I)), 
absence of all nodes with label ``B'' in a particular timestamp (Fig.~\ref{fig:didatic_local}(c-II)), and all nodes of the network active in the last two timestamps (Fig.~\ref{fig:didatic_local}(c-III)).}

In our case, LargeNetVis provides the TAM layout for the community being analyzed and also shows a line chart depicting the number of edges per timestamp in that community. \blue{Nodes are positioned according to their metadata, and nodes with the same metadata are positioned in the order they appear. We discuss the impact of using other strategies for node positioning in Sec.~\ref{limitation}.
When a square is hovered over in TAM, the system shows a tooltip with the respective timestamp, the node id, and node metadata. By clicking on a square, the corresponding node is highlighted in the node-link diagram.} We studied some \textit{alternatives} before adopting TAM, for example, the use of an animated node-link diagram instead of a static node-link diagram associated with the TAM. We chose the second option because timeline-based approaches are preferable for analysis that depend on more than two timestamps~\cite{SurveyDynamicVisualization}, \blue{for example during peak or valley analysis (T4).}

\subsection{\blue{Scalability}}
\label{scalability}

\blue{Some procedures in the system's workflow make LargeNetVis suitable for exploring large networks. First, node/edge sampling methods may be used as an initial attempt to reduce network data (Fig.~\ref{fig:workflow}(a)). 
Second, LargeNetVis detects communities throughout time and focuses on them (\ie, on their nodes and edges) to guide users to regions of interest in the network (Fig.~\ref{fig:workflow}(b)). Although this guidance is helpful when exploring large networks, the number of interesting communities may remain too large and additional guidance may be required; LargeNetVis efficiently addresses these cases by also providing community categorization (Fig.~\ref{fig:workflow}(b)). A potential side effect of focusing on the community structure is the discarding of inter-community edges and the consequent impact on the study of the overall structure and evolution of the whole network. Following previous work~\cite{CNO}, we consider this acceptable since inter-community edges are not as frequent as the intra-community ones~\cite{Fortunato2016} and nodes connected by them would create groups not as meaningful as the actual communities.}

\blue{To attenuate visual scalability issues in Global View depending on the number of timeslices being used and communities detected, a default zoom-to-fit is employed, so the representation of the smallest communities is reduced to a few pixels in the worst-case scenario.
Further efforts include from zoom-in/out, pan, and scrolling, to the selection of communities from categories of interest using the Taxonomy Matrix and the filtering of communities with a minimum desired number of nodes. Using a temporal network with 50,514 nodes and 108,132 edges (Twitter network~\cite{TwitterNetwork}), supp. material (Sec.~B) shows a case where Global View contains 2,033 communities and 100 timeslices.} 

\blue{The visual exploration of a community of interest (Fig.~\ref{fig:workflow}(c)) may also be impaired depending on its number of nodes and edges. To attenuate this scalability issue, besides spring-force node positioning and edge bundling, LargeNetVis shows by default a summarized node-link diagram when the number of nodes in the community is higher than a user-defined threshold (Sec.~\ref{systemLocalView}). In our supplemental material (Sec.~B), we show an example of a summarized representation of a community with 3,886 nodes and 5,035 edges. Free navigation between summarized and non-summarized diagrams is allowed and both views are coordinated between themselves and with TAM, thus optimizing the process of finding and analyzing nodes (or groups) of interest in all views. Finally, we have also analyzed the computational scalability of LargeNetVis by (i) collecting user feedback regarding the overall system's response time, and (ii) measuring the running times of the network manipulation procedures (Fig.~\ref{fig:workflow}(b)) for all networks and configurations discussed throughout Sections~\ref{usageScenario}~and~\ref{ux}. LargeNetVis is fast and its response time was well-received by our user study participants. We will provide details about this analysis in Sec.~\ref{results_ux}. }

\subsection{Implementation}

LargeNetVis was built with a client-server architecture and is available at \url{https://github.com/claudiodgl/LargeNetVis}. All views were implemented using Javascript and the D3 library~\cite{d3js}. The server side uses
Python and employs some popular frameworks and libraries, such as Flask~\cite{flask}, NetworkX~\cite{networkX_lib}, and others.

\section{Usage scenarios}
\blue{We present two usage scenarios to show the LargeNetVis capabilities.}

\label{usageScenario}

\subsection{MovieLens network}
\label{usageScenario_movielens}

The network we consider in this usage scenario (MovieLens network) was built from the \textit{MovieLens 1M dataset}~\cite{redesStream, movieLens}, which contains data gathered from the same-name rating website (\url{https://movielens.org/}). The network is composed of 9,940 nodes (3,900 users and 6,040 movies) and 1,000,209 edges (movie ratings on a 5-star scale) distributed in 1038 timestamps (days). It is a bipartite network where edges are defined as ``users rating movies''. We consider this network large because of its number of nodes and timestamps. Regardless of the number of timeslices tested in our experiments, all communities detected are born and die during the same timeslice (T3). The evolutionary behaviors growth/contraction and merge/split were not identified. We believe this occurs because there is probably no user rating the same movies more than once.

As stated in Sec.~\ref{systemOverview}, LargeNetVis allows users to freely explore communities directly from Global View, a resource that may lead to new insights and unexpected pattern discovery. The presented usage scenario begins with such an exploratory analysis (Fig.~\ref{fig:teaser}). We first filtered the MovieLens network to consider only negative ratings (\ie, with 1 or 2 stars, out of 5) for the top-6 most rated movie genres (action, comedy, documentary, drama, horror, thriller). Our focus was on those movies that belong to a single category among the mentioned ones, which resulted in 6,358 nodes and 47,809 edges distributed in 1,002 timestamps. Then, we adopted 10 timeslices to represent this observation period (value empirically chosen among the suggested range) and discarded communities with less than 5 nodes. 

With this network configuration, we quickly identified on Global View a timeslice (\#7) with more communities than the others (T1). This timeslice has 10 more communities than the timeslices with the least number of communities (\#2 and \#3, see Fig.~\ref{fig:teaser}(B)) and one more than the timeslices containing the second-highest number of communities (\#4 and \#5). We then became interested in exploring this additional community from timeslice \#7 (Fig.~\ref{fig:teaser}(F)). We clicked on this community's circle on Global View and the system informed us, through the Taxonomy Matrix cell highlighting, that this community is categorized as \textit{star} and \textit{sporadic/dispersed}, characteristics that represent 13 communities in the network (see Fig.~\ref{fig:teaser}(A)) (T2). We also noticed that this community contains an equal number of nodes and edges (33), and only 3\% of its timestamps have activity (Fig.~\ref{fig:teaser}(E)).

Moving further from global to local analysis, LargeNetVis shows the node-link diagram and TAM for any community selected on Global View. As we see in the node-link diagram from Fig.~\ref{fig:teaser}(C), our community of interest is composed of two individuals that rated several movies negatively. While one of these individuals rated movies from different genres (person P1), another rated only comedies (person P2). Note also that two movies were rated by both of them: \textit{House Arrest} (1996) and \textit{Life with Mikey} (1993). By looking at the TAM layout, we see another behavior shared by both individuals: except for two movies rated by P1 some days later (a drama and a comedy), each individual took only one day to rate all movies (Fig.~\ref{fig:teaser}(D)) (T1,T4).

\begin{figure}[t]
	\includegraphics[width=\linewidth]{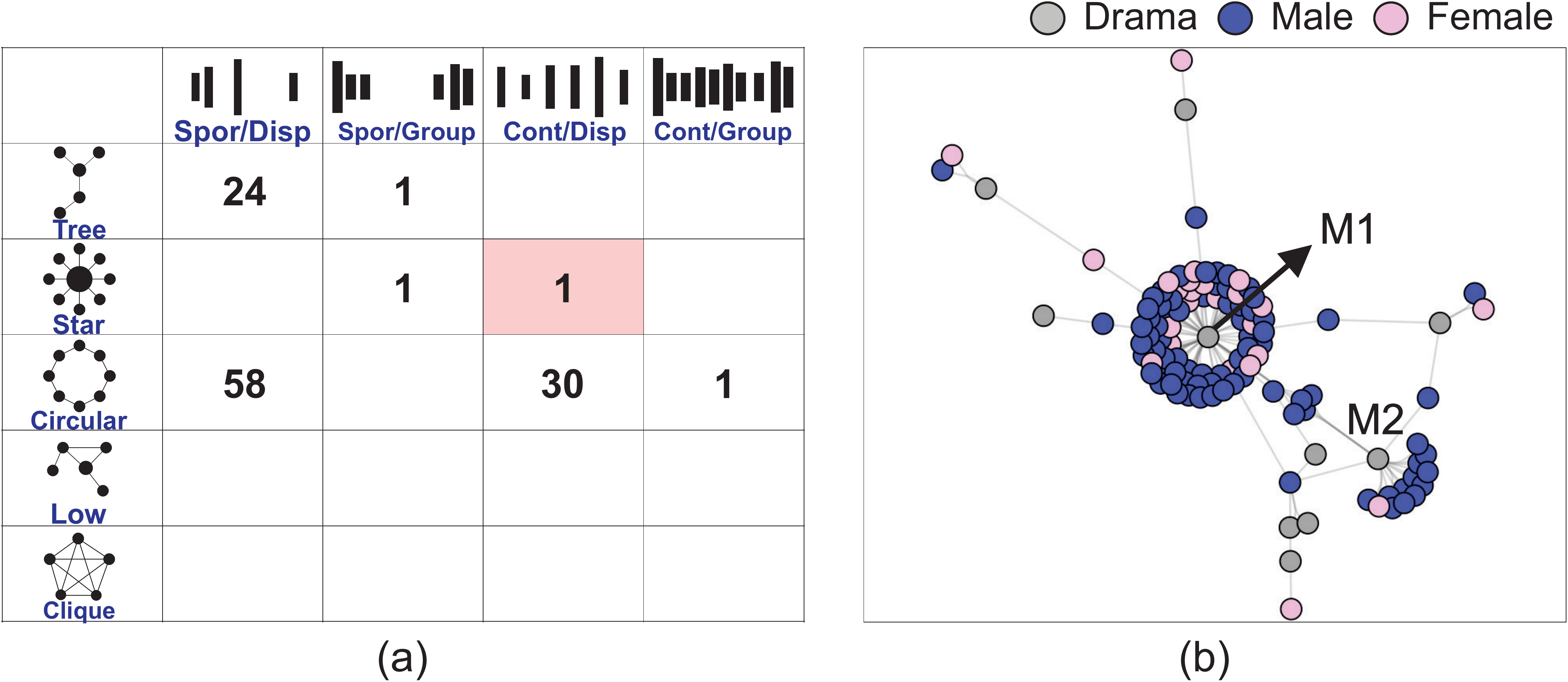}
	\centering
	\vspace{-0.7cm}
	\caption{\blue{The only \textit{star} and \textit{continuous/dispersed} community when considering drama movies positively rated. It was found thanks to the Tax. Matrix (a) and depicts two movies M1 and M2 rated by several people~(b).}}
	\label{fig:usage_scenario_analyse3}
	\vspace{-0.5cm}
\end{figure}

Fig.~\ref{fig:teaser}(C) illustrates a scenario where a single individual rates several movies; however, the inverse is also possible. By focusing the analysis on drama movies that were positively rated (\ie, with 4 or 5 stars, out of 5) and considering again 10 timeslices and communities with at least 5 nodes, we end up having 5,875 nodes and 72,695 edges distributed in 1,038 timestamps (days). The network under this configuration has only 1 community categorized as \textit{star} and \textit{continuous/dispersed}, which attracted our attention when looking at the Taxonomy Matrix \blue{(Fig.~\ref{fig:usage_scenario_analyse3}(a))} (T2). This community has 112 nodes (9 movies and 103 users) and contains two movies that were positively rated by a lot of people, mainly by males (Fig.~\ref{fig:usage_scenario_analyse3}(b)). The most rated one (M1) is \textit{One Flew Over the Cuckoo's Nest} (1975), winner of 5 Oscars and top-rated movie \#18 of all times on IMDb\footnote{\url{https://www.imdb.com/title/tt0073486/}. Accessed: 2022-03-22.}; the second movie (M2) is \textit{The Hurricane} (1999), nominated for an Oscar in 2000\footnote{\url{https://www.imdb.com/title/tt0174856/}. Accessed: 2022-03-22.}.

\subsection{\blue{Primary School network}}
\label{usage_scenario_primary}

The Primary School network~\cite{primarySchool} is composed of interactions recorded during two days between students and teachers of a primary school in Lyon, France. There are 242 nodes and 125,773 edges distributed in 5,846 timestamps (each comprising a 20-sec interval). The network considers five school grades (first to fifth) and two classes (A and B) per grade, in a total of 10 classes. Each class has an assigned teacher. We consider this network a large one due to its number of timestamps.

\begin{figure}[t]
	\includegraphics[width=\linewidth]{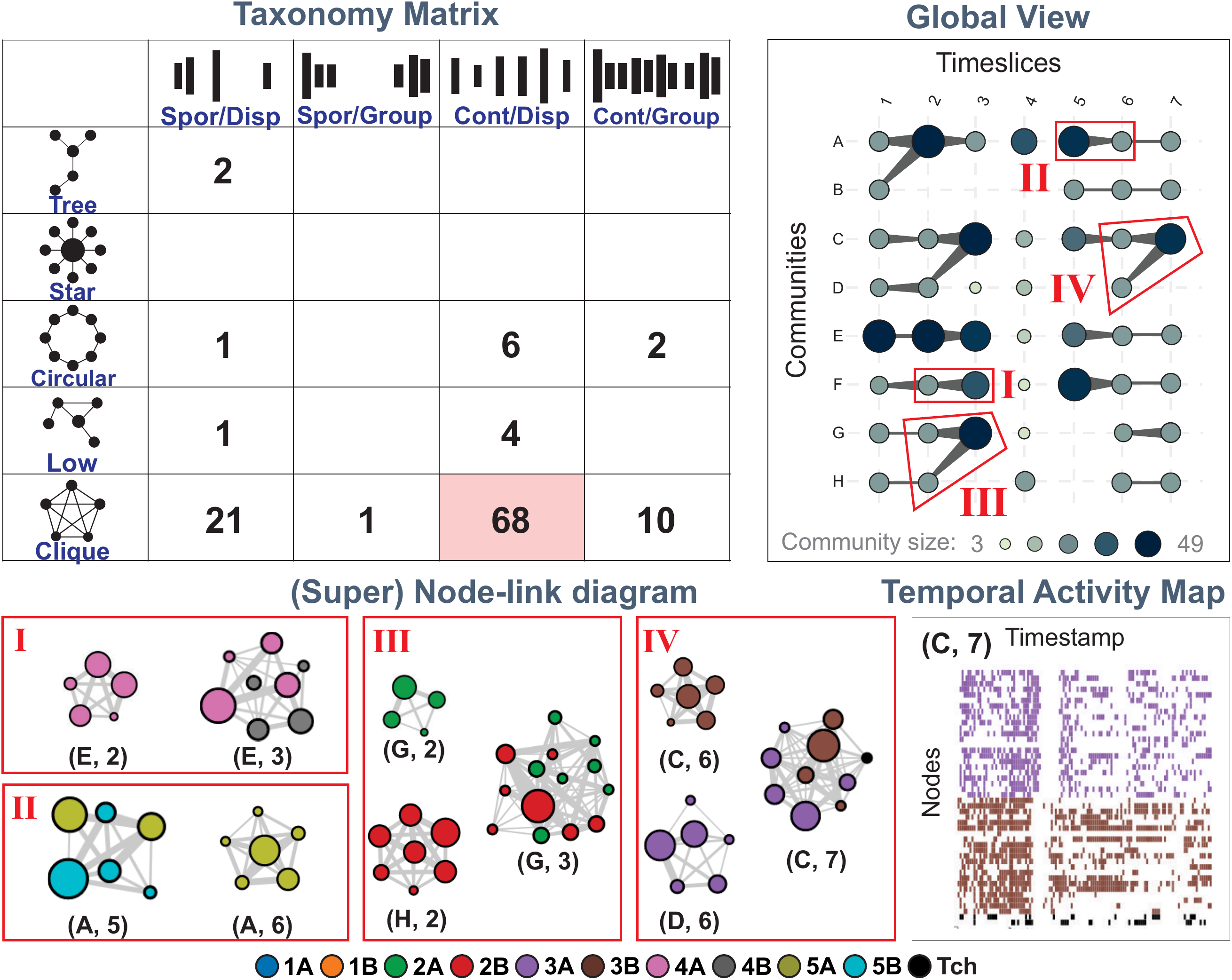}
	\centering
	\vspace{-0.6cm}
	\caption{\blue{Four evolutionary behaviors involving some of the 68 \textit{clique} and \textit{continuous/grouped} communities in the Primary School network: (I) a growth, (II) a contraction, (III, IV) two merges. The summarized node-link diagram is shown.}}
	\label{fig:primary_case_study}
	\vspace{-0.5cm}
\end{figure}

\blue{By adopting 26 timeslices and communities with at least three nodes to analyze this network, we see that most communities are categorized as \textit{clique} by the structural taxonomy (100 of 116) and that there are no \textit{stars} in the network (Fig.~\ref{fig:primary_case_study}, Tax. Matrix). Besides, there are 12 \textit{continuous/grouped} communities while \textit{sporadic/grouped} is the less observed temporal categorization, with only one community (T2). Note that the presence of communities with these characteristics is the opposite of what we observed in the MovieLens network (Fig.~\ref{fig:teaser}(A)), where there are no \textit{continuous/grouped} communities or  \textit{cliques} (MovieLens is bipartite, after all).}

\blue{When analyzing the first day of school (timeslices \#1 to \#7) using Global View, we see at least one growth (Fig.~\ref{fig:primary_case_study}(I)), one contraction (Fig.~\ref{fig:primary_case_study}(II)), and two merges (Fig.~\ref{fig:primary_case_study}(III-IV)) involving some of the 68  \textit{clique} and \textit{continuous/dispersed} communities (T1, T2, T3).
As shown in Fig.~\ref{fig:primary_case_study}, these four events occur between communities that represent same-grade classes (\eg, IV is a merge between classes 3A (purple) and 3B (brown)). This is expected since same-grade students interact more often among themselves than with students from other grades~\cite{primarySchool}. Finally, the merged community from IV is highly active over time except for a small time interval (T4) (see TAM in  Fig.~\ref{fig:primary_case_study}), a behavior that does not occur as frequently in MovieLens as in this network (recall Fig.~\ref{fig:teaser}(D)). This discrepancy may explain why the Primary School have most communities with \textit{continuous} activity while MovieLens have most communities having \textit{sporadic} activity (compare the Taxonomy Matrices from Fig.~\ref{fig:primary_case_study} and Fig.~\ref{fig:teaser}(A)).}

\section{User study}
\label{ux}

We also performed a user evaluation to collect feedback about the system's usefulness and usability, and ideas for improvement.

\subsection{Network data sets}

The participants of the user study employed LargeNetVis to analyze two different networks. The first one is the Primary School network presented in Sec.~\ref{usage_scenario_primary}. The second network, Sexual contacts~\cite{sexual_contacts}, describes sexual encounters between sex-workers (sellers) and clients (buyers). This network is composed of 12,157 nodes, 34,060 edges, and 1,000 timestamps (each timestamp comprises a 1-day interval). We consider it a large network mainly because of its number of nodes.

\subsection{Procedure}

Our study was conducted both online and remotely, with each participant answering the questions at different times and using their personal computers, so we did not have any control over the environment of the study. We provided each participant with a video containing instructions about the system's basic concepts and functionalities. The video was 7 minutes long and continuously available during the experiment.

\subsection{Online questionnaire}

The questionnaire was written in Brazilian Portuguese, in which all participants were fluent. We divided the questions in (i) background and experience; (ii) four basic questions to test the user's comprehension; (iii) six advanced questions using both networks (two multiple-choice and four open questions); (iv) Likert-scale-based questions to evaluate the user preference of the system, along with open questions to justify their choices; and (v) a mix of multiple-choice and open questions to collect the users' feedback about the system. The complete description of the questions and expected answers are available in the supplemental material (Sec.~C). This questionnaire structure was based on similar user studies evaluating layouts or systems~\cite{8440810, 8440832}.

The questions were divided aimed at evaluating the functionalities, layouts, and perceptions of the participants under different tasks. First, we assessed the comprehension of the basic functionalities of LargeNetVis using four basic multiple-choice questions about the Primary School (B1 -- B4). Then, we asked the participants to re-open the network under a different configuration (timeslices and filters), which led to different layouts used to answer the first three advanced questions (A1 -- A3), composed of one multiple-choice and two open questions. Finally, the last three questions (A4 -- A6, one multiple-choice and two open questions) are related to the Sexual contacts network. 
In the advanced questions, the first open question for each network (A1 and A4) was focused on guiding the participants to specific patterns, and the last questions (A3 and A6) encouraged them to perform a free exploration of the system.

\subsection{Participants}

The experiment recruited 14 participants, which participated voluntarily in the experiment. They are professors, postdocs, and graduate students with a background in Computer Science. We asked about their previous experience working in the Visualization and Network science fields: most have published articles or advised students in at least one of the fields. One participant also has extensive experience working with visualization in a number of international companies.
We also asked if they were aware of any visual difficulties, such as color blindness, but no one indicated limitations. \blue{They had no access to our usage scenarios.}

\subsection{Results}
\label{results_ux}

\myParagrapho{\blue{Participants' background.}} 
We divided the participants into non-overlapping groups based on their self-description of experience using a scale with None, Basic, Intermediate, and Advanced knowledge for each field (Visualization and Network science). We considered a participant as a \textit{Specialist} if he/she has advanced knowledge in both fields (we have 3 specialists). If he/she has advanced knowledge in only one field, we included him/her in the \textit{Advanced} group (3 individuals). Participants with an intermediate knowledge in at least one of the two fields were included in the \textit{Intermediate} group (8 individuals). These three groups were sufficient to categorize all 14 participants.

The participants spent 50m34s on average to answer the questionnaire and perform exploratory analyses. Considering the analyzed groups, the time spent answering the questionnaire increased based on the experience, with specialists spending almost one hour. We assume that participants with more knowledge of the fields spent more time assessing technical aspects of the layouts and analyzing their findings.

\myParagrapho{\blue{Questionnaire answers.}} 
All participants answered the four basic questions (B1 -- B4) correctly. Two of our advanced multi-choice questions (A2 and A5) aimed at validating the layouts. In question A2 (about Global View), we provided a 5-Likert scale asking the participants their agreement level on the ease of identifying communities that merge, with all participants providing positive answers (43\% agreed and 57\% strongly agreed). In question A5 (about the node-link diagram and TAM), they had to describe the structural pattern and temporal distribution of a particular community. All participants marked the correct answer. Having all correct answers to our multi-choice questions, we validated that the participants understood and correctly performed the intended tasks for all considered networks and configurations.

The open questions A1, A3, A4, and A6 were formulated to enable exploratory tasks. Questions A1 and A4 (one from each network) were designed to guide the participants in exploring specific patterns. A1 asked participants to indicate communities that suggest the occurrence of a lunch break in the primary school; we obtained 92\% of correct answers. Question A4 asked the participants to explore a given community using both the node-link diagram and TAM, and describe eventual patterns found. Again, 92\% of the participants indicated the expected answer, which was the relationship between one seller and several buyers, with connections spread over time.

Questions A3 and A6 asked participants to freely explore the layouts from both networks and describe relevant patterns eventually found, \blue{which should be different from those described in other questions. In this case, the participants were free to write their personal findings.} Our objective with these questions was not to evaluate correct or wrong answers, but to understand the overall perception of the users with the visualizations and the system's capability of enabling pattern identification. In question A3 (about the primary school), 72\% of the participants answered that they could identify at least one pattern not previously described in other questions. One pattern mentioned by 30\% of participants refers to the presence of the biggest network community on the second day of the network. This community comprises connections among students from six different grades and three teachers, which was very unusual for this network. \blue{We discuss this and other patterns mentioned by the participants in the supplemental material (Sec.~D).}

\begin{figure}[t]
	\includegraphics[width=1\linewidth]{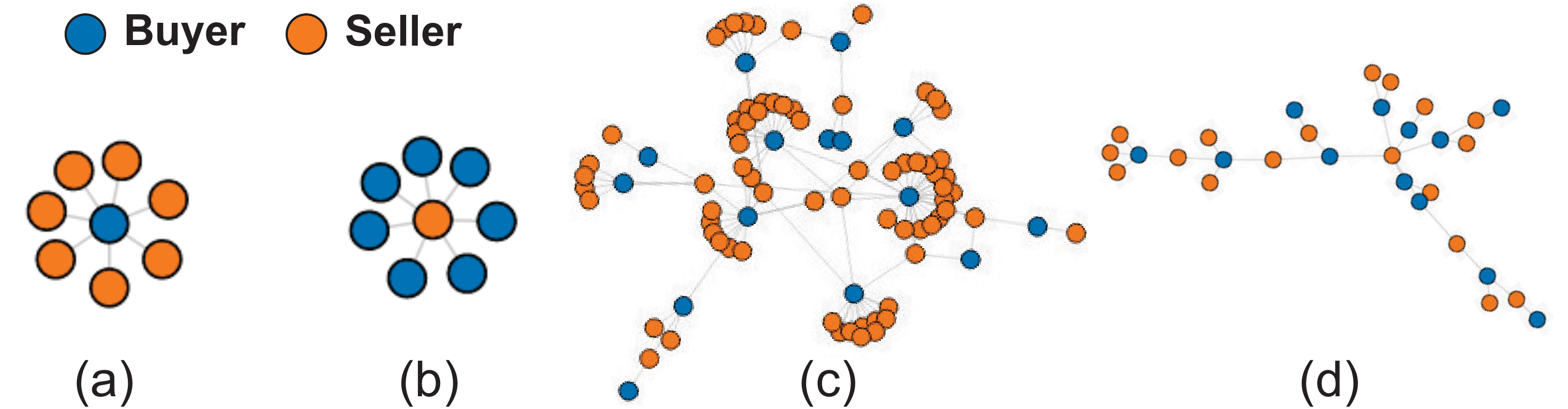}
	\centering
	\vspace{-0.6cm}
	\caption{Sexual contacts: some patterns identified through question A6.}
	\label{fig:exploratory_patterns_Sexual}
	\vspace{-0.5cm}
\end{figure}

In question A6 (about the Sexual contacts network), 57\% of the participants answered they could identify at least one pattern not previously described in other questions. The majority of the answers were related to the rarity of a seller's relationship with multiple buyers, for cases of small (Fig.~\ref{fig:exploratory_patterns_Sexual}(a-b)) and large communities (Fig.~\ref{fig:exploratory_patterns_Sexual}(c)). Also, one participant extended this analysis (Fig.~\ref{fig:exploratory_patterns_Sexual}(b)) by highlighting an interesting pattern in the TAM layout which shows that most encounters are 1-1 except for one involving 3 individuals on the same day. A participant also pointed out that \textit{``the proportion of star-type communities is much smaller than tree and circular, where inadequate protection can cause the spread of sexually transmitted infections''}. About the Tree type of structural pattern (Fig.~\ref{fig:exploratory_patterns_Sexual}(d)), a participant highlighted that \textit{``many large communities are of the Tree type structure, which shows that they are communities formed with few connections, in which a buyer can buy from multiple sellers, but there won't be many other buyers buying from these same sellers''}. This participant's observation is in agreement with this network's Assortativity value, which indicates a tendency of active buyers interact with less active sellers and vice-versa~\cite{sexual_contacts}.

\begin{figure}[t]
	\includegraphics[width=1\linewidth]{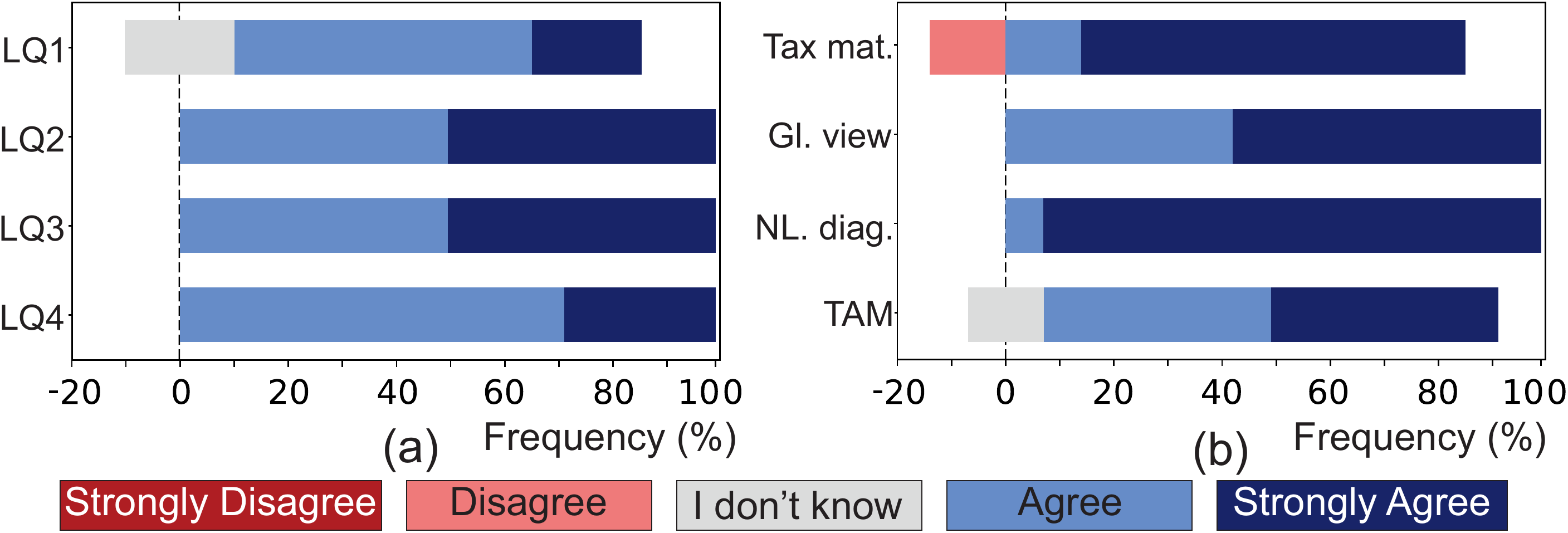}
	\centering
	\vspace{-0.6cm}
	\caption{Participants' answers related to (a) LargeNetVis and (b) individual layouts.  The bar length is the percentage of respondents who chose a specific Likert level. Questions from (a) concern the system's visual scalability (LQ1), speed (LQ2), usefulness (LQ3), and intuitiveness and ease of use (LQ4). Questions from (b) concern whether each layout is useful and helps in the network analysis.}
	\label{fig:likert_porc}
	\vspace{-0.5cm}
\end{figure}

\myParagrapho{\blue{LargeNetVis performance.}}
We used two 5-point Likert-scale questionnaires to measure the participants' preferences about LargeNetVis and specifically about each layout used (Fig.~\ref{fig:likert_porc}(a,b)). 
We evaluated four main aspects of the system: whether the system \blue{offers visual scalability (\ie, whether it worked well for both networks)} (LQ1), is fast (LQ2), is useful (LQ3), and is intuitive and easy to use (LQ4). 
Regarding LQ1, three participants (two with intermediate and one with advanced experiences) were not sure about the system's capability on this matter (Fig.~\ref{fig:likert_porc}(a), LQ1). The major concern was that the system was not tested using networks with 1 million nodes or so. In contrast, the specialists and other participants positively evaluated the system, indicating \textit{``a good balance between the number of elements in global and local views''}, and also that \textit{``it was possible to understand relationships in the same way in both networks''}. 

The other three aspects (Fig.~\ref{fig:likert_porc}(a), LQ2 -- LQ4) were positively evaluated by all participants. Some interesting aspects highlighted by the participants about LQ2 were that \textit{``responses to commands are immediate and instantaneous''} and \textit{``the interaction worked very fast, it just took a while to load the sexual network at first''}.
\blue{To obtain a quantitative assessment, we also measured the average running times for the workflow procedures depicted in Fig.~\ref{fig:workflow}(b), and we show our results in the supp. material (Sec.~E). The system spent a maximum of 12.66 sec. to run all of these procedures for any network and number of timeslices discussed throughout our usage scenarios and user study. Both the user feedback and quantitative evaluation demonstrate the well-received response time and computational scalability of LargeNetVis.}

\looseness=-1
Regarding the system's usefulness (Fig.~\ref{fig:likert_porc}(a), LQ3), participants claimed that the system \textit{``is very useful for dealing with unknown data in an exploration stage, such as observing the behavior of temporal data''}, and \textit{``this type of tool is vital for analyzing the evolution of communities in a temporal network''}. Regarding intuitiveness and ease of use (Fig.~\ref{fig:likert_porc}(a), LQ4), participants argued that \textit{``it is easy to use based on the graphical interface proposed and is quite intuitive for those who know networks or graphs''} and \textit{``I found it simple to use, being easy to switch between instants of time and follow the changes in relationships''}.

\myParagrapho{\blue{Feedback on individual layouts.}} 
Fig.~\ref{fig:likert_porc}(b) shows the participants' assessments of the quality of each layout, more specifically on whether the layout is useful and helped in the network analysis. The first evaluated layout was the Taxonomy Matrix, the only one to receive negative reviews. The critique of the two participants in this case was that they did not use this layout since the taxonomy information was also available under interaction in the tooltips shown in Global View. TAM also received two neutral assessments, justified by the participants who affirmed that \textit{``to completely understand TAM it is necessary to have a better knowledge of the network''}, and that \textit{``TAM needs better interaction techniques, such as the selection of a node in the node-link diagram and its automatic selection in the TAM layout''}. \blue{We implemented this feature after the user study.} Besides those neutral and negative assessments, participants considered that \textit{``the integration of all layouts facilitated the analysis and each brought a different aspect of the network being analyzed''}. Another participant even complemented this idea by explaining that \textit{``the Taxonomy Matrix allows a quick interpretation of communities, the node-link diagram is very useful to analyze the cases separately, and TAM adds the missing temporal information''}. Overall, the participants preferred Global View and the node-link diagram over the others.

Among the participants, 28\% mentioned other systems that allow similar analysis, such as Gephi, Graphviz, Stanford Network Analysis Platform (SNAP), and Python libraries (\eg, NetworkX). According to them, LargeNetVis is preferred because \textit{``it is already a specific and ready-to-use tool for network analysis''},\textit{ ``it doesn't mess up the graphs as much and it is more intuitive and easy to use''} and \textit{``it is a low-code tool; users only need to format their input data and the visualization is ready to use''}. On the other hand, disadvantages of our system include \textit{``the lack of flexibility in positioning views and exporting filtered data''} and \textit{``it could be possible to use the views as part of other systems''}.

We also asked the participants to evaluate the efficiency of the three taxonomies. In general, participants had very positive answers, claiming that \textit{``the taxonomies provide more objectively analyses, working as a filter of the network''} and also citing an example: \textit{``if I am interested in a certain structure, I discard the others and optimize time and efforts''}. A participant argued that the Taxonomy Matrix \textit{``shows to the user the pre-extracted and already categorized patterns, allowing less experienced users to analyze the data more directly and easily''}. 

\myParagrapho{Suggestions for improvement.}
Some participants suggested implementing edge weights to work with weighted networks, selecting multiple items in the layouts, improving the interaction between the node-link diagram and TAM \blue{(implemented after the user study)}, and others. The previous version of the system only exhibited network-level numerical information, and the most interesting suggestion we received in that sense was to optimize the numerical information panel by changing its information according to the granularity of analysis (network, community, or node-level information). 

\section{Discussion and Limitations}
\label{limitation}

Although LargeNetVis is a useful tool for exploring large temporal networks, some aspects may affect the analysis and pattern identification.

\myParagrapho{Number of nodes, edges, and timestamps.} \blue{Exploring networks with more than a few thousand nodes and timestamps (\ie, \textit{very} large networks) could lead to
a number of communities (or sizes) so elevated that the analysis would be impaired even considering LargeNetVis's current capabilities (\eg, the multilevel node-link diagram). 
We believe LargeNetVis is a encouraging first step towards the analysis of such very large networks, and we now intend to extend the system by incorporating more sophisticated filters (\eg, based on centrality values) and multilevel approaches in the other views (TAM and Global View).}

\myParagrapho{Node and community positioning.} 
\blue{As studied by previous work~\cite{Elzen2014,CNO,dynetvisSac,ForceDirectedStructural}, node positioning strongly affects overlaps, visual clutter, and pattern identification. As an example, identifying merges in Global View or interesting groups of nodes in the node-link diagram and TAM would be easier or harder depending on the adopted positioning. Even though the positioning algorithms already incorporated into the system have been sufficient for the tested cases, we now intend to include more sophisticated algorithms, \eg, \textit{Community-based node ordering}~\cite{CNO} for TAM and ad-hoc solutions specific for each type of structure. }

\myParagrapho{Multiple categorizations of the same community.} LargeNetVis's current version associates each community with a single category of the structural taxonomy. In real-world networks, however, the same community may contain more than one type of structure. One such example is the community shown in Fig.~\ref{fig:exploratory_patterns_Sexual}(c), which contains both \textit{star} and \textit{tree}. We now intend to adapt the system to consider these cases.

\myParagrapho{\blue{Network characteristics.}} \blue{Some taxonomies' categories may not be perceived depending on the network characteristics and dynamics. One such example is the absence of cliques in the bipartite network MovieLens (Sec.~\ref{usageScenario_movielens}), a type of community that would also not be identified in unipartite networks with low number of triangles. We have evaluated LargeNetVis using both unipartite and bipartite networks to show its usefulness for analyzing networks with different characteristics.}
\blue{LargeNetVis is designed for networks with discrete timestamps, for which timeslicing strategies are suitable~\cite{DynNoSlice}. Also, the system adopts uniform (equally distributed) timeslices instead of non-uniform and data driven ones. In future work, we intend to investigate whether data driven timeslices would improve our results.}

\myParagrapho{Information loss.}
LargeNetVis offers a set of sampling methods that can be used to discard nodes and edges from the analysis. \blue{It also focuses on the network communities, discarding inter-community edges and making it infeasible the analysis of the overall structure and evolution of the whole network (recall Sec.~\ref{scalability}).} Although evaluating the performance of sampling methods is outside the scope of our study, we now intend to measure how much information is lost\,---\,and how relevant they are\,---\,when inter-community edges are ignored.

\myParagrapho{\blue{Visual Improvements.}}
\blue{Based on reviewers' comments and participants' feedback, we added some new features to the system after the user study (adaptive numerical information panel, varying link thickness in Global View, multilevel node-link diagram, edge bundling, and coordination between node-link and TAM). These features were not used by the participants and did not directly affect the reported results. We now intend to improve multiple community selections in Global View and thoroughly evaluate all these new features.}

\section{Conclusion}

In this paper, we presented LargeNetVis, a web-based visual analytics system that assists experts and practitioners in analyzing small and large temporal networks, \ie, networks varying from a few nodes and timestamps to a few thousand of these elements.
LargeNetVis enhances the analysis and pattern identification by leveraging the network community structure and three taxonomies, each accounting for a particular and relevant type of pattern (structural, temporal, and evolutionary). 
Given a temporal network and its node metadata, the system first divides the network into timeslices of equal length. Then, it detects network communities inside each timeslice. These communities are then categorized using three taxonomies from the literature. Finally, users can explore communities with characteristics of interest and their nodes and edges through the four linked views offered by the system, two focused on global level analysis and two focused on a local level. We validated LargeNetVis's usefulness and effectiveness through two usage scenarios and a user study with 14 participants.


 \acknowledgments{
This work was supported by grants \#2020/10049-0, \#2020/07200-9, and \#2016/17078-0 from S\~ao Paulo Research Foundation (FAPESP), grant \#E-26/201.424/2021 from Carlos Chagas Filho Foundation for Research Support of Rio de Janeiro State (FAPERJ), CNPq, CAPES, and by the School of Applied Mathematics at Fundação Getulio Vargas (FGV). The authors thank Luis R. Pereira for providing an optimized version of his source-code~\cite{PEREIRA2021100904}.
}

\bibliographystyle{abbrv-doi}

\bibliography{paper}
 
\end{document}


\maketitle

\appendix
\section{\blue{Link length reduction algorithm}}

\blue{Algorithm~\ref{algoritmo} depicts the link length reduction strategy we use in Global View.}

\begin{algorithm*}

\caption{\blue{Link length reduction strategy}}
\label{algoritmo}

\begin{algorithmic}[1]

\Procedure{}{}
    \If{first grid column (timeslice)} 
        \State Insert communities throughout the grid column in the order they appear
    \EndIf
    
    \For{each remaining grid column $X$}
        \State $previousColumn$ $\gets$ grid column X-1
        
        \For{each ``from'' community ($fromC$) in $previousColumn$}
            \For{each corresponding ``to'' community ($toC$) in X} \Comment{there are two ``to'' communities in splits}
                \State $toPosition$ $\gets$ position in $X$ such that EuclideanDistance(position, $fromC$) is minimized
                \State Put $toC$ in $toPosition$ 
                \State Draw link between $fromC$ and $toC$
            \EndFor
        \EndFor
        
        \State Insert communities not linked so far in the available column positions in the order they appear
        
        \For{each ``to'' community in X that is result of a merge} \Comment{let's minimize merge-related diagonals}
            \State $oldFromPosition \gets$ position in $previousColumn$ where its fartest ``from'' community is located
		    \State $newFromPosition \gets$ position in $previousColumn$ where the non-linked community closest to the closest ``from'' is located
		    \State Switch communities in $oldPosition$ and $newPosition$
        \EndFor
    \EndFor
    
\EndProcedure
\end{algorithmic}
\label{algoritmo}

\end{algorithm*}

\section{Visual scalability}

\blue{Fig.~\ref{fig:teaser} shows a screenshot of the LargeNetVis system when analyzing the Twitter network~\cite{twitter}, which depicts retweets mentioning a famous Brazilian newspaper (\textit{Folha de S\~ao Paulo}). There are 50,514 nodes (Twitter users) and  108,132 edges distributed in 224 timestamps, each comprising a 1-h interval. An edge is created when a user's tweet that mentions the newspaper is retweeted by another user --- both users then become nodes in the network. When using 100 timeslices for analysis, the network is decomposed into 2,033 communities (Fig.~\ref{fig:teaser}(B)) that are simultaneously shown in Global View due to the default zoom-to-fit (Fig.~\ref{fig:teaser}(C)). There are 331 communities categorized as \textit{Star} and \textit{Continuous/Grouped} (Fig.~\ref{fig:teaser}(A)), including the largest community in this network (whose tooltip is being exhibited in Global View), with 3,886 nodes and 5,035 edges. Fig.~\ref{fig:teaser}(D) shows the most summarized version of its node-link diagram. Note that this representation perfectly matches this community's structural categorization (\textit{Star}).}

\section{Questionnaire for the user study}

The questionnaire was originally written in Brazilian Portuguese, in which all participants were fluent. The questions were translated to English in this document. For the basic and advanced questions, correct or expected answers are marked in bold.

\subsection*{Background and experience}

\begin{enumerate}
    \item Are you aware of any visual difficulties you may have?
    \item What area is your education in (e.g., computer science, statistics)?
    \item What is your most relevant academic title/function? Choose one option: (i) I'm an undergraduate student; (ii) I'm pursuing my master's degree; (iii) I'm a PhD student/candidate; (iv) I'm a postdoctoral researcher; (v) I'm a professor.
    \item What is your degree of prior knowledge in the Information Visualization field? Choose one option: None, Basic, Intermediate, and Advanced knowledge.
    \item What is your degree of prior knowledge in the Network Science field? Choose one option: None, Basic, Intermediate, and Advanced knowledge.
    \item Briefly explain your experience with the fields mentioned above (Visualization and Network Science).
\end{enumerate}

\subsection*{Four basic questions (B1 -- B4)}

Given the Primary School network with default filters, answer:

\begin{description}
\item[B1] Considering the ``Taxonomy Matrix'' with the predefined parameters (``X: Temporal tax.'' and ``Y: Structural tax.''), select the two communities categorized as \textit{Clique} and \textit{Continuous and Grouped}. Answer in which positions they are found in the ``Global View'': (i) (A, 4) and (B, 5); (ii) (F, 12) and (D, 15); (iii) (B, 14) and (C, 16); \textbf{(iv) (F, 13) and (D, 14)}; (v) (G, 3) and (B, 4).

\item[B2] In ``Global View'', select the community (A, 13) and analyze its corresponding ``Node-link diagram''. This community is composed by the majority of students from school class: (i) 5A; (ii) 2B; \textbf{(iii) 5B}; (iv) 1A; (v) 3A.

\item[B3] The community (A, 13) in ``Global View'' is categorized by the three taxonomies as: \textbf{(i) Cont/Disp, Clique and Death}; (ii) Cont/Disp, Star and Birth; (iii) Spor/Group, Circular and Merge; (iv) Spor/Disp, Tree and Clique; (v) Cont/Disp, Tree and Low.

\item[B4] The community (E, 13) in ``Global View'' is composed of teacher and students from class 4B. In the next timeslice (14), what happens to this community? \textbf{(i) Other students join the community, mainly students from class 4A}; (ii) It splits into two other communities; (iii) A group of teachers joins the community; (iv) Students from class 1A join the community; (v) Students from classes 2A and 2B join the community.

\end{description}

\subsection*{Six advanced questions (A1 -- A6)}

Given the Primary School network with 35 timeslices and communities with at least 15 nodes, answer:

\begin{description}

\item[A1] The Primary School network consists of two days, divided by a night period where there is no activity. In addition, there is a lunch break on each day where students from at least three classes meet and interact with each other. Please describe two communities that illustrate expected lunchtime behavior. Answer using the format (row, column), which corresponds to the position of the community in ``Global View''. \textbf{Expected answer: Almost any community between timeslices 5 -- 6 and 30 -- 32, e.g., (A, 32) and (D, 30). \blue{Note that the main challenge here is to find one or two timeslices that is(are) related to the lunch break(s). There are only five correct timeslices among the 35 possibilities.}}

\item[A2] Between timeslices 7 and 8 of the ``Global View'', classes 1A (community (F, 7)) and 5B (community (G, 7)) merge and form a single community (F, 8). You may zoom-in to visualize this behavior better, if necessary. How do you evaluate the statement ``It is easy to identify communities that merge''? Choose one option: (i) Strongly disagree; (ii) Disagree; (iii) I don't know; (iv) Agree; (v) Strongly agree.

\item[A3] Based on the Primary School network, do you think it is possible to find other patterns in addition to those mentioned above? If so, please describe one that you consider relevant and mention which part of the visualization helped you find it.

\end{description}

Given the Sexual contacts network with 18 timeslices and communities with at least 5 nodes, answer:

\begin{description}

\item[A4] In the ``Taxonomy Matrix'', select ``X: Structural tax.'' and ``Y: Structural tax.'' and click ``Change''. Select the 16 communities categorized as Star. A star community is mostly composed of a central node that interacts with the other nodes in the group. Among the 16 communities of this type highlighted in the ``Global View'', how do you interpret the behavior of the community (F, 9)? Please analyze the node-link diagram and the Temporal Activity Map (TAM) to support your answer. \textbf{The expected answer is related to the relationship between one seller and several buyers, with connections spread over time.}

\item[A5] Regarding the community (T, 12), choose the option where all statements are true (if any): (i) It is a Tree. It has only buyers. There are three timestamps with interactions; (ii) It is a Star. It has 3 buyers and 3 sellers. There is a timestamp with five iterations; (iii) It is a Star. It has 1 buyer and 3 sellers. There is at least one interaction at each timestamp; \textbf{(iv) It is a Tree. It has 3 buyers and 3 sellers. There is a timestamp where two buyers interact with the same seller;} (v) None of the above. 

\item[A6] Based on the Sexual contacts network, do you think it is possible to find other patterns in addition to those mentioned above? If so, please describe one that you consider relevant and mention which part of the visualization helped you find it.

\end{description}

\subsection*{Basic/Advanced  questions and intended layouts}

Table~\ref{tab:question_layout} associates the basic and advanced questions with the visual components they are related to. The basic questions (B1 -- B4) and the first advanced question (A1) rely on the use of two visual components to support the participants' answers. In general, Global View was the layout that could be used for all tasks since it represents the overview of the system and leads to other views. Also, since questions A3 and A6 asked the participants to freely explore the system and find new patterns, any layout could potentially be used in this discovery. The second advanced question (A2) requires the analysis of a single component. The last four advanced questions (A3 -- A6) depend on the analysis of multiple views.

\begin{table}[t]
\centering
\caption{Basic/Advanced questions and the visual components they are related to.}
\label{tab:question_layout}
\resizebox{0.49\textwidth}{!}{%
\begin{tabular}{lllllllllll}
\hline
                     & B1                        & B2                        & B3                        & B4                        & A1                        & A2                        & A3                        & A4                        & A5                        & A6                        \\ \hline
Tax mat.       & \checkmark &                           & \checkmark &                           &                           &                           &                         \checkmark   & \checkmark & \checkmark & \checkmark \\
Gl. view           & \checkmark & \checkmark & \checkmark & \checkmark & \checkmark &     \checkmark                       &                        \checkmark    & \checkmark & \checkmark & \checkmark \\
NL. diag.     &                           & \checkmark &                           &            \checkmark               & \checkmark &  & \checkmark & \checkmark & \checkmark & \checkmark \\
TAM &                           &                           &                           &                           &  &                           &                        \checkmark    & \checkmark & \checkmark & \checkmark \\ \hline
\end{tabular}%
}
\end{table}

\subsection*{Questions to compare LargeNetVis with other systems}

\begin{enumerate}

\item Are you able to do this type of analysis with another system?

\item What other system(s) that allow similar analysis do you know?

\item What are the advantages of LargeNetVis compared to the system(s) you are familiar with?

\item What are the disadvantages of LargeNetVis compared to the system(s) you are familiar with?

\end{enumerate}

\subsection*{Likert-scale questions (LQ1 -- LQ8)}

We used two 5-point Likert-scale questionnaires to assess the participants' preferences about LargeNetVis (LQ1 -- LQ4) and the provided visual components (LQ5 -- LQ8). For each question below, the participants should choose between (i) Strongly disagree; (ii) Disagree; (iii) I don't know; (iv) Agree; (v) Strongly agree.

\begin{description}

\item[LQ1] Do you agree with the statement ``LargeNetVis offers visual scalability (i.e., it works well for both networks)''?

\item[LQ2] Do you agree with the statement ``LargeNetVis is fast (i.e., the provided interactions work in a satisfactory time)''?

\item[LQ3] Do you agree with the statement ``LargeNetVis is useful''?

\item[LQ4] Do you agree with the statement ``LargeNetVis is intuitive and easy to use''?

\end{description}

\begin{description}

\item[LQ5] Do you agree with the statement ``The `Taxonomy Matrix' is useful and helped when analyzing the networks''?

\item[LQ6] Do you agree with the statement ``The `Global View' is useful and helped when analyzing the networks''?

\item[LQ7] Do you agree with the statement ``The `Node-link diagram' is useful and helped when analyzing the networks''?

\item[LQ8] Do you agree with the statement ``The `Temporal Activity Map (TAM)' is useful and helped when analyzing the networks''?

\end{description}

\subsection*{Questions to collect the users' feedback about the system}

\begin{enumerate}

\item Do you agree with the statement ``The three taxonomies and how LargeNetVis uses them contribute to a more efficient network analysis''? Leave a comment.

    \item In your opinion, what are the most useful visual aids offered by LargeNetVis? Why?
    
    \item What other visual aids do you think could be useful if incorporated into the LargeNetVis system?
    
    \item Do you have any final comments?
    
\end{enumerate}

\section{Primary School - Patterns from question A3}

\begin{figure}[t]
	\includegraphics[width=1\linewidth]{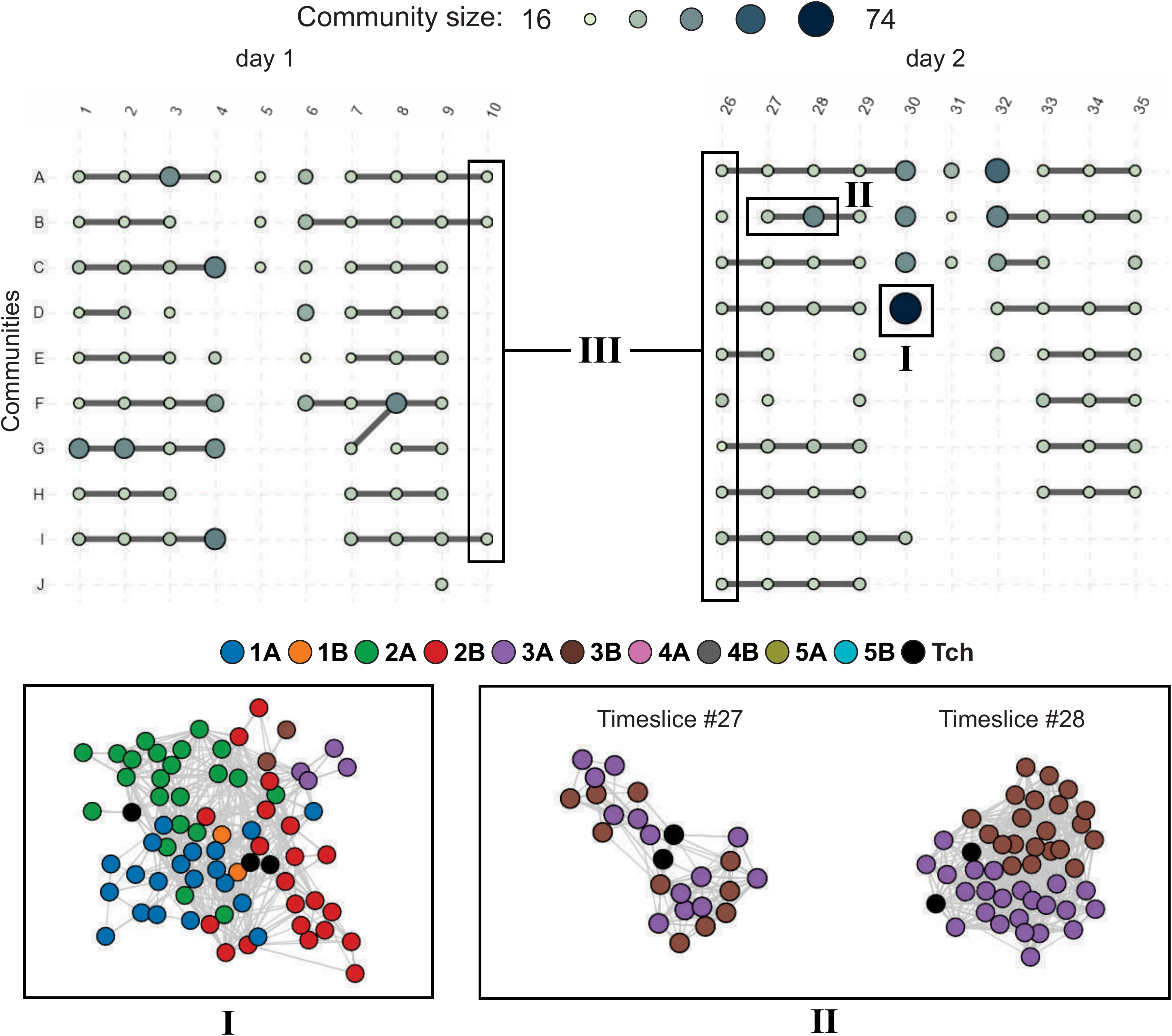}
	\centering
	\vspace{-0.6cm}
	\caption{Primary school: some patterns identified through question A3. (I) Presence of the biggest community in the second day of the network. (II) A community growing between timeslices 27 and 28. (III) Spor/Group temporal categorization matches the beginning and end of class periods. \blue{Varying link thickness to map growths and contractions were not available in the version of Global View used in the user study. This feature was implemented based on reviewers' suggestions and feedback from the participants.}}
	\label{fig:exploratory_patterns_School}
	\vspace{-0.3cm}
\end{figure}

Fig.~\ref{fig:exploratory_patterns_School} illustrates some interesting patterns described by participants for the Primary School network (question A3). Fig.~\ref{fig:exploratory_patterns_School}(I) shows the presence of the biggest network community in the second day of the network. Its existence was pointed out by 30\% of participants. This community comprises connections among students from six different grades and three teachers, which was very unusual for this network (see first node-link diagram in Fig.~\ref{fig:exploratory_patterns_School}). Some assumptions created by the participants were that this pattern represents \textit{``some type of special event''}, or \textit{``an increase of the lunch-break duration''}. Fig.~\ref{fig:exploratory_patterns_School}(II) shows a community growing between timeslices 27 and 28, which represents an increase in the number of students from the 3rd grade (see the corresponding node-link diagram). At last, two participants indicated that the Sporadic and Grouped temporal categorization guided them to find new patterns on the Global view, which matched the beginning and end of class periods (highlighted in Fig.~\ref{fig:exploratory_patterns_School}(III)).

---

\begin{table}[b]
\centering
\caption{\blue{Average running times (in seconds) for the networks and numbers of timeslices considered in our usage scenario and user study. We also show the average times for the Twitter network mentioned in Sec.~B. \#C and  \#T refer to the number of communities detected and number of timeslices considered, respectively. Columns 1-5 refer to the average running time spent to: (1) suggest a range of suitable numbers of timeslices, (2) perform the community detection, (3-5) categorize all communities according to the structural, temporal, and evolution taxonomies, respectively.}}
\label{tab:execution_time}
\resizebox{0.5\textwidth}{!}{%
\begin{tabular}{ccccccccc}
\hline
\textbf{Network} & \textbf{\#C} & \textbf{\#T} & \textbf{1} & \textbf{2} & \textbf{3} & \textbf{4} & \textbf{5} & \textbf{\begin{tabular}[c]{@{}c@{}}Total \\ (sec)\end{tabular}} \\ \hline
Primary          & 81           & 16           & 4.33       & 5.45       & 0.31       & 2.55       & 0.02       & 12.66                                                           \\
Primary          & 115          & 26           & 4.59       & 5.73       & 0.26       & 1.52       & 0.03       & 12.13                                                           \\
Primary          & 147          & 35           & 4.11       & 5.63       & 0.23       & 1.07       & 0.03       & 11.07                                                           \\
Sexual           & 669          & 18           & 0,6        & 5,73       & 0.43       & 0.21       & 0.69       & 7.66                                                            \\
movie Neg.   & 154          & 10           & 0.7        & 6.78       & 0.37       & 0.24       & 0.17       & 8.26                                                            \\
movie Pos.   & 116          & 10           & 0.56       & 7.45       & 0.48       & 0.35       & 0.14       & 8.98                                                            \\ \hline
Twitter   & 2,033          & 100           & 0.12       & 12.35       & 1.46       & 0.47       & 2.82       & 17.22                                                            \\ \hline
\end{tabular}%
}
\end{table}

\section{Running Times}

\blue{Table~\ref{tab:execution_time} depicts the average running time of 10 executions for every procedure of the LargeNetVis workflow depicted in Fig. 3(b) of the main paper, i.e., the average running time needed to (1) suggest a range of suitable numbers of timeslices, (2) perform the community detection, (3-5) categorize all communities according to the structural, temporal, and evolution taxonomies, respectively. The experiments were performed on a personal computer with Intel(R) Core(TM) i7-7700K CPU @ 4.20GHz, 16 GB RAM, video card GeForce GTX 1070 8GB, and Windows 10.}

\blue{Note that the system spent a maximum of 12.66 sec. to run all of these procedures for any network and number of timeslices discussed throughout our usage scenarios and user study. LargeNetVis is still very fast (17.22 sec. on average) when analyzing the Twitter network (discussed in Sec.~B). Recall that this network has 50,514 nodes and 108,132 edges. It was decomposed into 100 timeslices and 2,033 communities, numbers many times greater than those from the other networks (Table~\ref{tab:execution_time}).
Every user interaction (zoom, selection, change of colors, etc.) lasts 1 second in the worst case for any network. Both the user feedback and our quantitative analysis demonstrate the well-received response time and computational scalability of LargeNetVis.}

